\documentclass[aps,prb,superscriptaddress,reprint,floatfix,twocolumn]{revtex4-1}
\usepackage{graphicx,amssymb,amsmath,bm,dcolumn}
\usepackage[caption=false]{subfig}
\usepackage{natbib}
\usepackage{color}
\usepackage{threeparttable}
\usepackage[bottom]{footmisc}
\usepackage{comment}
\usepackage{xfrac}
\usepackage{amsmath,amssymb}
\usepackage{appendix}
\usepackage{enumitem}

\definecolor{dukeblue}{rgb}{0.0, 0.0, 0.65}

\newcommand{\bl}[1]{\color{dukeblue}{#1}}

\begin{document}
\title{Detecting continuous gravitational waves with superfluid $^4$He}

\author{S. Singh}
\altaffiliation[Present address: ]{Department of Physics, Williams College, Williamstown, Massachusetts 01267, USA.}
\affiliation{Department of Physics, College of Optical Sciences and B2 Institute, University of Arizona, Tucson, Arizona 85721, USA}
\affiliation{ITAMP, Harvard-Smithsonian Center for Astrophysics, Cambridge, MA 02138, USA}
\email{swati.singh@williams.edu}

\author{L.A. De Lorenzo}
%
\affiliation{Applied Physics, California Institute of Technology, Pasadena, CA 91125 USA}

\author{I. Pikovski}
\affiliation{ITAMP, Harvard-Smithsonian Center for Astrophysics, Cambridge, MA 02138, USA}
\affiliation{Department of Physics, Harvard University, Cambridge, MA 02138, USA}

\author{K.C. Schwab}
\affiliation{Applied Physics, California Institute of Technology, Pasadena, CA 91125 USA}

\email{schwab@caltech.edu}
\date{\today}

\begin{abstract}
Direct detection of gravitational waves is opening a new window onto our universe. Here, we study the sensitivity to continuous-wave strain fields of a kg-scale optomechanical system formed by the acoustic motion of superfluid helium-4 parametrically coupled to a superconducting microwave cavity. This narrowband detection scheme can operate at very high $Q$-factors, while the resonant frequency is tunable through pressurization of the helium in the 0.1-1.5 kHz range. The detector can therefore be tuned to a variety of astrophysical sources and can remain sensitive to a particular source over a long period of time. For reasonable experimental parameters, we find that strain fields on the order of $h\sim 10^{-23} /\sqrt{\rm Hz}$ are detectable. We show that the proposed system can significantly improve the limits on gravitational wave strain from nearby pulsars within a few months of integration time.
\end{abstract}

\maketitle

\section{Introduction}
The recent detection of gravitational waves (GW) marks the beginning of gravitational wave astronomy \cite{PhysRevLett.116.061102}. The first direct detection confirmed the existence of gravitational waves emitted from a relativistic inspiral and merger of two large black holes, at a distance of $400M $ parsecs (pc). Indirect evidence for gravitational radiation was previously attained by the careful observation since 1974 of the decay of the orbit of the neutron star binary system PSR B1913+16 at a distance of 6.4 kpc, which agrees with the predictions from general relativity to better than 1\% \cite{Weisberg2005}. In this paper, we discuss the potential to use a novel superfluid-based optomechanical system as a tunable detector of narrow-band gravitational wave sources, which is well suited for probing nearby pulsars at a distance of less than 10kpc. As we discuss below, in the frequency range exceeding $\sim$500 Hz, this novel scheme has the potential to reach sensitivities comparable to Advanced LIGO.

\begin{figure}[ht]
\begin{center}
\includegraphics[width=3.2in]{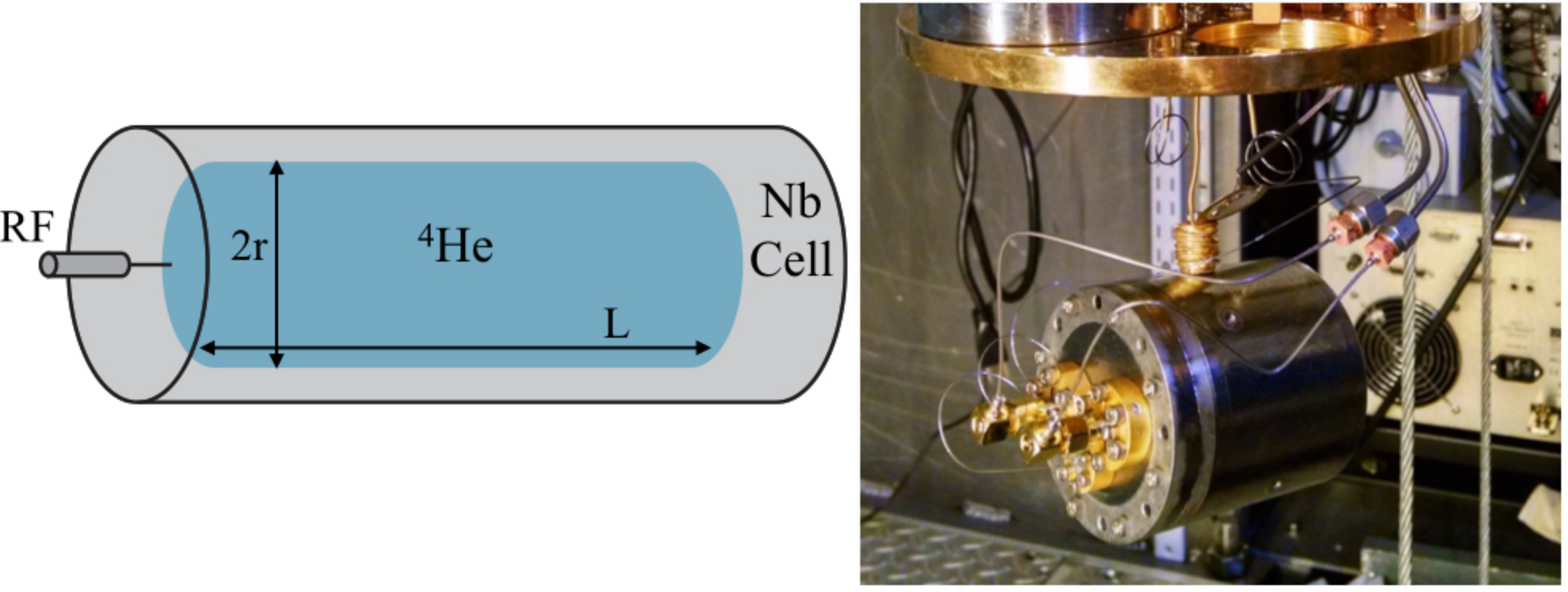}
\caption{{\bf Left:} Schematic of the proposed gravity wave sensor based on acoustic modes of superfluid helium. Two cylindrical geometries considered here are Gen1 (radius $a= 11$cm, length $L= 50$cm, mass $M= 2.7$kg) and Gen2 ($a= 11$cm, $L=3$m, $M= 16$kg). {\bf Right:} Prototype of the detector with $a= 1.8$cm, $L=4$cm, $M=6$g and resonant frequency 10 kHz.}
\label{fig:HeDet}
\end{center}
\end{figure}

The GW detector under consideration is formed by high-$Q$ acoustic modes of superfluid helium parametrically coupled to a microwave cavity mode in order to detect small elastic strains. This setup was initially studied in Ref. \cite{DeLorenzo2014}, and is shown in fig. \ref{fig:HeDet}. The helium detector effectively acts as a Weber bar antenna \cite{Weber1960} for gravitational waves, but with two important differences. Firstly, the $Q/T$-factor of the helium is expected to be much larger than that of metals, where $Q$ is the acoustic quality factor, and $T$ is the mode temperature. Secondly, the acoustic resonance frequency can be changed by up to 50\% by pressurization of helium without affecting the damping rate, making the detector both narrowband and tunable.

The power spectrum of gravitational waves is expected to be extremely broad and is estimated to range from $10^{-16}$ to $10^3$ Hz \cite{Shutz1989, Cutler2002, Riles2013} for known sources. Ground-based optical interferometers (such as LIGO, Virgo, GEO, TAMA) allow for a broad-band search for gravitational waves in the frequency range 10 Hz - 1 kHz. These detectors are expected to be predominantly sensitive to the chirped, transient, GW impulse resulting from the last moments of coalescing binaries involving compact objects (black holes(BH) and/or neutron stars (NS)) \cite{Abadie2010}. Space-based interferometric detectors can in principle be sensitive to lower frequency gravitational waves, as they are not limited by seismic noise \cite{Vitale2014}. 
 
Unlike broadband impulse sources, rapidly rotating compact objects such as pulsars are expected to generate highly coherent, continuous wave gravitational wave signals due to the off-axis rotating mass, with frequencies spanning from $\sim$1 kHz for millisecond pulsars (MSPs) in binaries, to $1$ Hz for very old pulsars \cite{Shklovskii1970, Ostriker1969, Ferrari1969, Melosh1969, Shutz1989}. Given the unknown mass distribution of the pulsar, one can only estimate the strain field here at earth. However, several mechanisms give upper bounds to the strength of gravitational waves on earth. One such limit is the ``spin down limit", which is given by the observed spin-down rate of the pulsar, and the assumption that all of the rotational kinetic energy which is lost is in the form of gravitational waves \cite{Abbott2008PRD}. Another limit is given by the yield strength of the material which makes up the neutron star, and how much strain the crust can sustain before breaking apart due to centripetal forces \cite{Owen2005}. The presence of strong magnetic fields indicate a potential mechanism for producing and sustaining such strains due to deformation of the neutron star \cite{Pitkin2011}. However, without knowing the strength and direction of the internal magnetic fields in a pulsar, it is difficult to estimate a lower limit on the size of gravitational wave signal. The measurement of gravitational waves from pulsars would therefore give us crucial information about the interior of neutron stars. 

Since pulsars should emit continuous and coherent gravity waves at specific and known frequencies, we can use a narrowband detector and integrate the signal for long times, averaging away the incoherent detector noise. We show that for reasonable parameters, the superfluid helium detector can approach strain sensitivities of $1-5\times 10^{-23}/\sqrt{\rm Hz}$ at around 1 kHz, depending on the size and $Q$ factor of the detector. Pulsar frequencies are observed to vary slightly due to random glitches $\Delta f / f \sim 10^{-6}-10^{-11}$ (older, millisecond pulsars being more stable) \cite{LyneBook2006}, and due to the motion of the earth around the sun and resulting doppler frequency shifts. The tunability of the acoustic resonance will be essential to track these shifts during long detection integration times.  Simultaneous monitoring of the targeted pulsar electromagnetically can facilitate the required precision frequency tracking.  The frequency agility can also allow for using the same acoustic resonator to look for signals from multiple pulsars in a similar frequency range.

Recent measurements with LIGO and Virgo have unsuccessfully searched for the signals from 179 pulsars and have limited the strain field $h\lesssim 10^{-25}$ for most pulsars after nearly a year of integration time\cite{Aasi2014}. 
In a parallel development, hundreds of new pulsars have been discovered in the last few years by analyzing Gamma-Ray sources observed by the Fermi Large Area Telescope (Fermi-LAT), some less than 0.5 kpc from earth \cite{Fermi1Catalog2010, Fermi2Catalog2013}. Together, these developments are signaling a promising path towards gravitational wave astronomy of pulsars.

This paper is organized as follows. We start with an overview of continuous gravitational waves from pulsars to get an estimate for the strains produced on earth in Section \ref{sec: AstroSources}. We then describe the superfluid helium detector and show how it functions as a detector for gravitational waves in Section \ref{Sec:HeliumDet}. In Section \ref{Sec:Noise_Hmin}, we provide the detection system requirements. We then compare this detector with other functional gravity wave detectors, and show the key fundamental differences between these detectors and our proposed detector in Section \ref{Sec:Compare}. Finally, we conclude with a brief summary of the key features of this detector and outlook in Section \ref{Sec:Conclusion}. A review of the relevant concepts and derivations are relegated to the appendices for the interested reader.

\section{Sources of continuous gravitational waves}
\label{sec: AstroSources}
The generation of gravitational waves can be studied by considering the linearized Einstein equations in the presence of matter \cite{MisnerBook1973}. The computations are similar to the analogous case in electromagnetism \cite{Schutz1984}, see Appendix A for details. However, in the absence of gravitational dipoles, a quadrupole moment $Q_{ij}$ is necessary to source gravitational waves. The emitted power of gravitational waves is found to be \cite{CarrollBook2004}
\begin{equation}\label{eq:power}
P = \frac{G}{5 c^5} \langle \dddot{Q}_{ij} \dddot{Q}^{ij} \rangle ,
\end{equation}
i.e. it depends on the third time derivative of the quadrupole moment of the system, where $Q_{ij}:=\rho\int_{\rm body} x_i x_j dV$ for a body of density $\rho$. 

 In the far-field limit where size of the source ($GM/c^2$) $\ll$ wavelength of gravity wave ($c/\omega$) $\ll$ distance to detector ($d$), the gravitational metric perturbation becomes
\begin{equation}
h_{ij}=\frac{2G}{c^4 d} \ddot{Q}_{ij},
\end{equation}
where $h$ is the gravitational perturbation tensor in transverse-traceless gauge. Since $G/c^4 \sim 10^{-44} {\rm N\ s^4/kg^2}$, one needs events with relativistic changes in mass quadrupole moment to have a measurable source of gravitational radiation on earth. As an estimate, if all the observed slowdown of the Crab pulsar was converted into gravitational radiation, the power would correspond to $P \sim 4.5 \times 10^{31}$ W ($10^5$ times the electromagnetic radiation power from the sun) \cite{Abbott2008}. However, at a distance 2 kpc away from the pulsar (distance to earth), the power flux is $10^{-9} {\rm W/m}^2$ and the metric perturbation is $h\sim 10^{-24}$.  Even though the power flux is macroscopic and easily detectable in other forms (acoustic, electromagnetic, etc.), the resulting strain is very small due to the remarkably high impedance of space-time.  This is at the heart of the difficulty with laboratory detection of gravity waves.

The several astrophysical candidates for gravitational waves considered so far can be broadly classified into three categories: stochastic background, broadband impulses, and continuous sources \cite{Ju2000, Adhikari2014}. It is estimated that there will be a broadband background of gravitational waves from the expansion of the early universe. Furthermore, there is a low frequency stochastic background due to gravitational waves emitted by masses moving in the galaxies.  Impulse sources could stem from supernovae or mergers of compact objects. The latter is the primary source being searched for by most ground based detectors, and was recently observed by the LIGO detectors \cite{PhysRevLett.116.061102}. Lastly, continuous gravitational waves can be expected from stellar binaries (albeit at very low frequencies), or from pulsars. We now discuss the generation of gravitational waves from asymmetric pulsars and the limits on the signal set on earth.

Estimates of gravitational radiation from pulsars is an active area of theoretical research that goes back to early observations of pulsars\cite{Melosh1969}. The mechanism for gravitational wave generation is assumed to be an asymmetric mass distribution. Several mechanisms are proposed for the deviation from axial symmetry in mass distribution, for example magnetic deformations, star quakes or instabilities due to gravitational or viscous effects \cite{Lasky2015,LyneBook2006}. However, due to the unknown equation of state, there is significant variability in estimates of mass asymmetry and thus gravitational wave strain from pulsars. Null results from measurements of GW strain from pulsars have already put limits on the equation of state \cite{Owen2005}. 

Assuming that the emission of gravitational waves is a contributing mechanism towards the observed slowdown of pulsars enables us to set upper bounds on the GW metric strain here on earth. We now estimate the gravitational wave perturbation strain from measured spin-down rates and briefly discuss the validity of this limit. We then present relevant numbers for a few millisecond pulsars (MSPs) of interest for our detector. Details about these derivations and typical parameters for other pulsars of interest are presented in Appendix B.

For an ellipsoidal pulsar rotating about the $z$-axis with frequency $\omega_p$, the two polarizations of $h$ are given by
\begin{eqnarray}
\label{eq:hmax}
h_+&=&-\frac{4G}{c^4 d} \epsilon I_{zz}\omega_p^2\cos{2\omega_p t},\\
h_\times&=&\frac{4G}{c^4 d} \epsilon I_{zz}\omega_p^2\sin{2\omega_p t},
\end{eqnarray}
where $I_{zz}$ is the moment of inertia along the $z$-axis and $\epsilon$ characterizes the mass quadrupole ellipticity ($\epsilon = (Q_{xx}-Q_{yy})/I_{zz}$). Thus, the strain changes at twice the rotation frequency of the pulsar. The energy flux for a continuous GW of polarization $A$ from a pulsar source is given by
\begin{equation}
s^A = \frac{c^3}{16\pi G} {\overline {\dot{h}_A \left( t \right)^2} },
\end{equation}
where $A\in \{+, \times\}$ and the bar indicates time-averaging.

Typical neutron stars have mass 1-1.5 $M_{\odot}$ (where $M_{\odot}=2 \times 10^{30}$ kg is the solar mass) and have a radius of around 10 km. Using these values, the moment of inertia amounts to $10^{38}$ kg-m$^2$, the estimate used in previous GW searches, c.f. Ref. \cite{Aasi2014}. The ellipticity parameter is estimated by assuming that the observed slow-down rate ($\dot\omega_p$) of a pulsar is entirely due to emission of gravitational waves. This estimate is then used to compute the upper-limit estimate for gravitation perturbation strain known as the spin down strain,
\begin{equation} 
\label{eq:hsdown}
h_{sd}=-\frac{4G}{c^4 d} \epsilon I_{zz}\omega_p^2=\sqrt{\frac{5GI_{zz}\dot\omega_p}{2c^3d^2\omega_p}}.
\end{equation}

From Eq. \ref{eq:hsdown}, it is clear that given the values of distance ($d$), rotational frequency ($\omega_p$) and spin-down frequency ($\dot{\omega}_p$) from astronomical observations, we can put limits on gravitational wave strains due to pulsars. 

While $h_{sd}$  is a useful first-principles upper limit, it over-estimates the strength of gravitational waves, particularly from young pulsars that are highly active electromagnetically. This has already been confirmed by braking index measurements \cite{Manchester2005, Palomba2000} and the negative results from recent GW detector data \cite{Aasi2014}. However, long-lived and stable millisecond pulsars ($\dot{\omega_p/2\pi <10^{-14}} {\rm Hz\ s^{-1}}$) such as the ones considered here have been proposed as likely sources of continuous gravity waves \cite{New1995}. The frequency stability and relatively low magnetic field indicate that unlike young pulsars like Crab and Vega, the dominant spin down mechanism in MSPs is more likely to be quadrupolar gravitational radiation.

We can also set limits on GW generation mechanism by considering specific models of the interior of neutron stars, as discussed in ref. \cite{Lasky2015}, and reviewed briefly in Appendix B. Assuming standard nuclear matter and breaking strain for elastic forces, the ellipticity sustained can be limited to less than $6\times 10^{-7}$, irrespective of the physics leading to deformations \cite{Ushomirsky2000,Owen2005}. This can also be used to evaluate the GW strain amplitude limit $h_{\epsilon n}$. Since the strain limits $h_{sd}$ and $h_{\epsilon n}$ come from different physics (conservation of angular momentum and balancing forces in stellar interior), we use the lower of the two as the upper limit for metric strain.

{\renewcommand{\arraystretch}{1.4} 
\begin{widetext}
\begin{center}
\begin{table}[ht]
\label{tab:MSPtable}
\begin{tabular}{|c |c| c| c| c| c| c|c|c|}
\hline
Pulsar & $\omega_p/2\pi$ & $f_{GW} $ (Hz) & $\dot{\omega_p}/2\pi$  (Hz s$^{-1}$) & $d$ (kpc) &$h_{sd}$ & $h_{\epsilon n}$ & $h_0^{95\%}$ & $h_{\rm He,1}^{95\%}$[l,m,n]  \\
\hline\hline

J0034-0534 & 532.71 & 1065.43 & $ -1.5 \times 10^{-16}$ & $0.5\pm .1$ & $2.7 \times 10^{-27}$ &$ 1.4 \times 10^{-24}$ &$1.8 \times 10^{-25}$ & $1.1 \times 10^{-26}$ [020]\\
\hline

J1301+0833$^*$ & 542.38 & 1084.76 & $ -3.1 \times 10^{-15}$ & $0.7\pm .1$ & $2.8 \times 10^{-27} $ & $ 1.1 \times 10^{-24}$ & $1.1 \times 10^{-25} $ & $1.0 \times 10^{-26}$ [020]\\
\hline

J1747-4036$^*$ & 609.76 & 1219.51 & $ -4.9 \times 10^{-15}$ & $3.4\pm0.8$ & $6.7 \times 10^{-28} $ & $2.8 \times10^{-25}$ & no data  & $8.8 \times 10^{-27}$ [020]\\
\hline

J1748-2446O & 596.44 & 1192.87 & $ -9.4 \times 10^{-15}$ & $5.9\pm .5$ & $5.4 \times 10^{-28} $ & $ 1.5 \times 10^{-25}$ & $2.6 \times 10^{-25} $ & $9.1\times 10^{-27}$ [020]\\
\hline

J1748-2446P & 578.50 & 1157 & $ -8.7 \times 10^{-14}$ & $5.9\pm .5$ & $1.7 \times 10^{-27} $ & $ 1.4 \times 10^{-25}$ & $1.6 \times 10^{-25} $ & $9.5 \times 10^{-27}$ [020]\\
\hline

J1748-2446ad & 716.36 & 1432.7 & $ -1.7 \times 10^{-14}$ & $5.9\pm .5$ & $6.7 \times 10^{-28} $ & $ 2.2 \times 10^{-25}$ & $1.8 \times 10^{-25} $ & $1.3 \times 10^{-26}$ [201]\\
\hline

J1810+1744$^*$ & 601.41 & 1202.82 & $ -1.6 \times 10^{-15}$ & $2.0\pm .3$ & $6.6 \times 10^{-28} $ & $ 4.6 \times 10^{-25}$ & $1.8 \times 10^{-25} $ & $9.0 \times 10^{-27}$ [020]\\
\hline

J1843-1113 & 541.81 & 1083.62 & $ -2.8 \times 10^{-15}$ & $1.7\pm .2$ & $1.1 \times 10^{-27} $ & $ 4.4 \times 10^{-25}$ & $1.1 \times 10^{-25} $ & $1.0\times 10^{-26}$ [020]\\
\hline

J1902-5105$^*$ & 574.71 & 1149.43 & $ -3.0 \times 10^{-15}$ & $1.2\pm .2$ & $1.5 \times 10^{-27} $ &$ 7.0 \times 10^{-25}$ & no data & $9.6\times 10^{-27}$ [020]\\
\hline

J1939+2134 & 641.93 & 1283.86 & $ -4.3 \times 10^{-14}$ & $3.6\pm .3$ & $5.8 \times 10^{-28} $ &$ 2.9 \times 10^{-25}$ &$1.3 \times 10^{-25} $ & $8.1\times 10^{-27}$ [020]\\
\hline

J1959+2048 & 622.12 & 1244.24 & $ -4.4 \times 10^{-15}$ & $2.5\pm .5$ & $8.6 \times 10^{-28} $ & $ 3.9 \times 10^{-25}$ & $1.5 \times 10^{-25} $ & $8.5\times 10^{-27}$ [020]\\
\hline

\end{tabular}
\caption{Table of millisecond Pulsars with rotation frequency greater than 500 Hz: $\omega_p$ is the rotational frequency, $f_{GW} =\omega_p/\pi$ is the frequency of gravitational waves, $\dot{\omega}_p$ is the measured spin down rate, and $d$ is the distance to the pulsar in kilo-parsecs, $h_{sd}$ and $h_{en}$ are the spin-down and elastic strain limits. These values are compared to the strain limit set by recent continuous GW surveys by interferometric detectors $h_0^{95\%}$ \cite{Aasi2014}, and the strain limit for two identical Gen1 helium resonant detectors (with an integration time of 250 days) as shown in Eq. \ref{eq:Hedet_hmin}. Pulsars indicated by superscript $^*$ were discovered by the Fermi gamma ray telescope. Pulsars J1747-4036 and J1902+5105 were only discovered in 2012 \cite{Kerr2012}, and were not included in the LIGO+VIRGO analysis presented in Ref. \cite{Aasi2014}}

\end{table}
\end{center}
\end{widetext}

Table \ref{tab:MSPtable} details parameters for pulsars of interest with rotational frequency higher than 500 Hz, along with current limitations on GW strain from LIGO+VIRGO collaboration. Theoretical estimates of metric strain assuming spin-down limit, and elastic crust breakdown limit on ellipticity ($\epsilon = 6\times 10^{-7}$) from Ref.\cite{Owen2005} are also given. Table \ref{tab:MSPtable} also gives the strain estimate set by the helium detector outlined in figure \ref{fig:HeDet} (Gen1) that we will discuss in detail in the following sections. Several of these pulsars were discovered recently by analyzing gamma-ray sources from Fermi-LAT. The number of known fast spinning pulsars is expected to grow significantly as more sources are discovered and analyzed. Furthermore, there is growing evidence that the GeV excess emission in our galactic center is in fact due to hundreds of unresolved MSPs and not from dark-matter annihilation \cite{Bartels2016, Lee2016}. 

Since the strain due to gravity waves from pulsars is expected to be very small but coherent, one needs to integrate the signal from these detectors for a long time (typically several days, the last result being a compilation of $\sim 250$ days of integration over multiple detectors \cite{Aasi2014} ). Also, in order to rule out noise we need to detect a gravitational wave signal from at least two different detectors. Strain sensitivity also improves as $\sqrt{N_d}$, where $N_d$ is the number of detectors \cite{Dupuis2005}. There has been computationally intensive analysis of many days of data from broadband detectors like LIGO to search for such gravitational wave signals. However, all such searches have so far been unsuccessful, although they have improved the upper bound on the emitted wave amplitudes. These upper bounds in turn constrain the equation of state of exotic neutron stars. In the following, we outline the proposal for a simple, low-cost, narrowband detector for these gravitational waves based on a superfluid helium optomechanical system. Being relatively simple and economical, superfluid helium detectors can also be set up in multiple locations to improve overall detection sensitivity. 

As a precursor to subsequent discussions, we present the central result of our work in figure \ref{fig:hminPulsars},  showing the limits set by different detectors for ten MSPs of interest from Table \ref{tab:MSPtable}. Along with the spin-down limit and the strain limit set up the previous LIGO measurement \cite{Aasi2014}, we also show the limits set up two different geometries of helium detectors that we discuss in detail in Sections \ref{Sec:HeliumDet} and \ref{Sec:Noise_Hmin}. Since we are detecting a continuous GW signal, the strain sensitivity improves with integration time. Here, we have assumed that the resonance frequency of the same acoustic mode can be tuned by up to 200 Hz without changing the $Q$-factor (of $10^{11}$), thereby resonantly targeting each of these pulsars with the same detector.

\begin{figure}[ht]
\begin{center}
\includegraphics[width=3.5in]{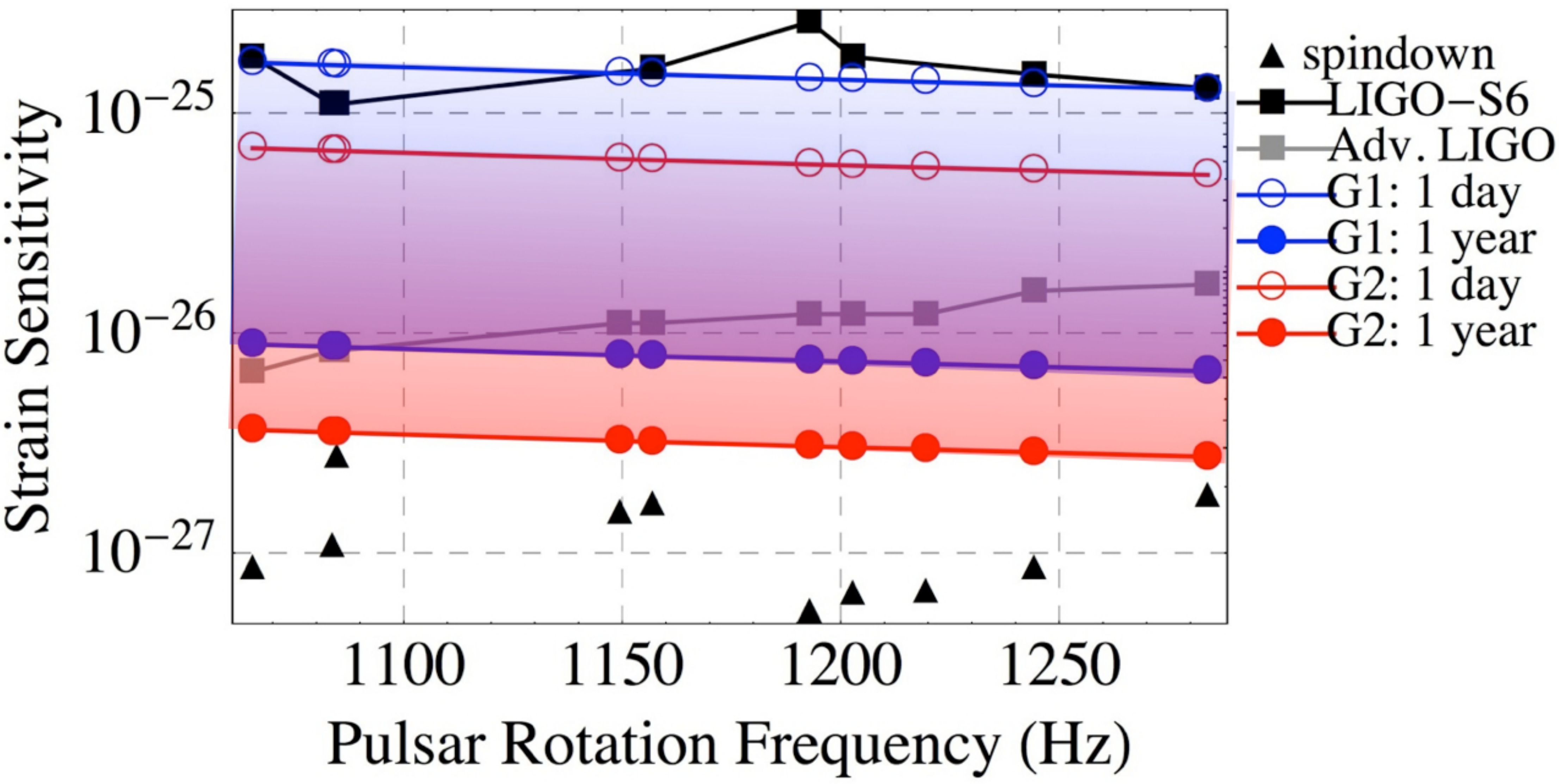}
\caption{Strain sensitivity of various detectors for 10 MSPs of interest for the helium detector versus measurement time for two helium detectors with same sensitivity operating simultaneously, assuming a bath temperature of 5mK, $(l,m,n)=(0,2,0)$, and $Q-$factor of $10^{11}$ for both Gen1(mass=2.66 kg, in blue) and Gen2(mass= 15.9 kg, in red) detectors. We also show the limits set by three interferometric sensors operating at LIGO-S6 sensitivity for 250 days \cite{Aasi2014} and at Advanced LIGO design sensitivity for one year\cite{Abbott2016ALIGOnum}. As seen in the figure, the current limit on pulsar GW strains can be surpassed within a few days of integration time for Gen1, and in under a day for Gen2. Also shown is the spin-down limit for these pulsars.}
\label{fig:hminPulsars}
\end{center}
\end{figure}
\section{Superfluid helium gravity detector}
\label{Sec:HeliumDet}

The gravitational wave strain detector we propose is a resonant mass detector formed by acoustic modes of superfluid helium in a cavity parametrically coupled to a microwaves in a superconducting resonator. For the purpose of our calculations, we will treat the superfluid as an elastic medium with zero dissipation.  At the temperatures we expect to operate this detector, $T<10$mK, the normal fluid fraction $\rho_n$ is expected to be $\rho_n/\rho_0<10^{-8}$, where $\rho_0$ is the total density of the fluid \cite{Donnelly1998}.  For temperatures below $T<100$mK, the dissipation of audio frequency acoustic waves is expected and found to be dominated by a three-phonon process, falling off as $T^{-4}$. 

An elastic body (with dimensions $\ll\lambda_{GW}$) in a gravitational field will undergo deformation due to changes in space-time as a gravity wave passes by. For distances far away from the source of radiation, the space-time perturbation acts like an external tidal force \cite{Hirakawa1973}, as also discussed in Appendix A. The equation of motion for the displacement field ${\bf u}\left(r,t \right)$ of an elastic body is given by \cite{LandauBook1986}
\begin{equation}
\label{eq:elasticbody}
\rho \frac{\partial^2 {\bf u}}{\partial t^2}-\mu_L \nabla ^2  {\bf u} -(\lambda_L+\mu_L)\nabla (\nabla.{\bf u})=\frac{1}{2}\rho \ \ddot{\bf h} {\bf x},
\end{equation}
where $\rho$ is the density, $\lambda_L,\mu_L$ are the Lam\'e coefficients for the elastic body and $\ddot{\bf h} {\bf x}$ is the effective amplitude of the wave for a particular orientation of the detector that exerts an effective tidal force on the detector. 

This acoustic deformation can be broken into its eigenmodes ${\bf u}\left({\bf r},t \right)=\sum_n\xi_n\left( t\right){\bf w}_n\left( {\bf r} \right)$.  For this analysis, we assume our acoustic antenna is in a single eigenmode of frequency $\omega_m$, thus dropping index $n$. In this analysis, we have used the notation where ${\bf w}_n\left( {\bf r} \right)$ is a dimensionless spatial mode function with unit amplitude, and the actual amplitude of the displacement field is in $\xi(t)$.

Rigid boundary walls and absence of viscosity enables us to describe the acoustic modes accurately via a simple wave equation as opposed to Navier-Stokes equations typically used to describe fluid flow. The spatial modes are obtained by solving the acoustic equations of motion \cite{KinslerBook2000}. For elastic deformations in enclosed spaces, the change of pressure $p({\bf r})$ is described by
\begin{equation}
\nabla^2 p - \frac{1}{c_s^2} \frac{\partial^2 p}{\partial t^2} =0
\label{eq:pressure}
\end{equation}
with the speed of sound in the material (here helium) being $c_s$. The particle velocity $\bf{v} = \dot{\bf{u}}$ is related to pressure via $ \partial \bf{v}/\partial t =  -  \nabla p/\rho$. Thus each vector component of the velocity $\bf{v}$ also satisfies the same wave equation as the pressure, but the components are not independent of each other. The full solution can be equivalently expressed in terms of the Helmholtz potential for the velocity, ${\bf v}=\nabla \Phi({\bf r})$. In terms of the potential, the acoustic pressure becomes $p= -\rho \partial \Phi / \partial t $, and the potential satisfies the same wave equation
\begin{equation}
\nabla^2 \Phi - \frac{1}{c_s^2} \frac{\partial^2 \Phi}{\partial t^2} =0
\label{eq:Helm}
\end{equation}
As before, the time dependence can be explicitly separated via $\Phi \rightarrow \Phi({\bf r}) \xi(t)$. For cylindrical symmetry the solution for the spatial part of the potential is
\begin{equation}
\Phi(r,\theta,z) = J_m(k_{m}(n) r) \cos(m \theta) \cos\left( k_z(l) (z+ \frac{L}{2})\right),
\label{eq:potential}
\end{equation}
where the wavevectors are found from the rigid boundary conditions $\partial \Phi/ \partial z = 0$ at $z = \pm L/2$ and $\partial \Phi/ \partial r = 0$ at r=a, such that $k_z(l) = l \pi/L$ with $l=0,1,2 ...$ and $k_{m}(n)$ follows from the n roots of $J'_m(k_{m}(n) a)=0$. Having the solution for the potential, one can obtain the velocity vector field, and thus the spatial modes, via ${\bf w}(r,\theta,z)=\nabla \Phi(r,\theta,z)/|{\bf w}_{\rm max}|$, where $|{\bf w}_{\rm max}|$ is the maximum value of $\nabla \Phi(r,\theta,z)$. These acoustic modes of helium in a superconducting cavity were experimentally studied by some of the authors in Ref. \cite{DeLorenzo2014}. We found these modes to be well-modeled {\bl by} this theory and to have extremely high $Q$-factors ($Q>10^{8}$) at 45 mK.

\begin{figure}[ht]
\begin{center}
\includegraphics[width=3.3in]{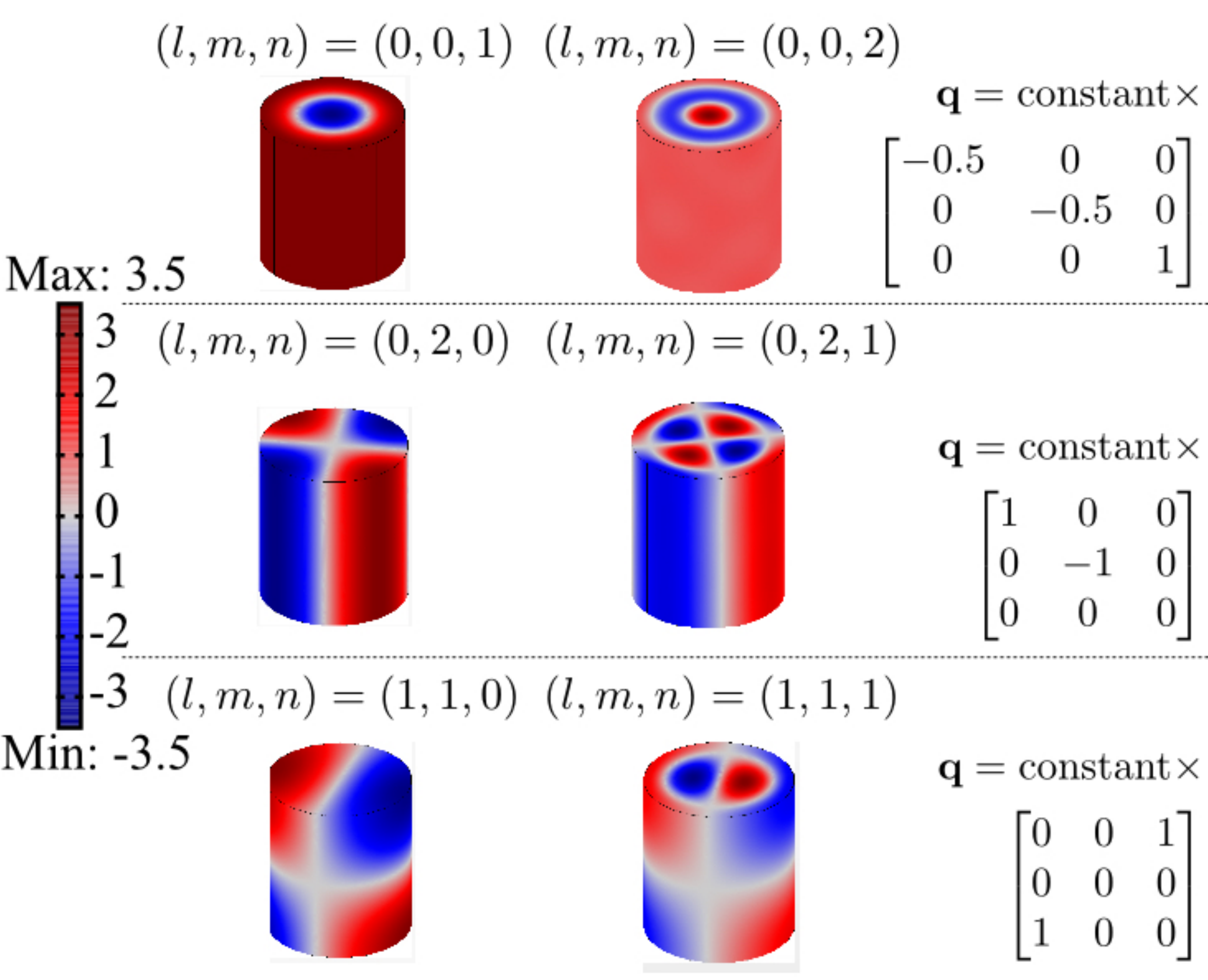}
\caption{ The first few pressure modes with non-zero quadrupolar tensors for the cylindrical cavity. While the form of quadrupolar tensor (shown on the right) is similar for many modes, the constant could be different for each acoustic mode. }
\label{fig:HeModes}
\end{center}
\end{figure}

For the purposes of this paper, we will simply add the finite linear dissipation to the acoustic resonance, parameterized as a finite $Q$. For a damped acoustic resonator, eq. \ref{eq:elasticbody} can be simplified to show that  the displacement field $\xi(t)$ satisfies the equation of motion
\begin{equation}
\mu \left(\ddot{\xi }+\frac{\omega_m}{Q_{He}}\dot\xi+\omega_m^2\xi \right) = \frac{1}{4}\sum_{ij} \ddot{h}_{ij}q_{ij},
\label{eq:EOM}
\end{equation}
where $Q_{He}$ is the $Q$-factor associated with the acoustic mode, $\mu$ is the reduced mass for the particular eigenmode,
\begin{equation}
\mu = \int \rho {\bf w}^2 dV,
\end{equation}
and $q_{ij}$ is the dynamic part of the quadrupole moment, 
\begin{equation}
q_{ij} = \int{\rho\left( w_i x_j+x_i w_j - \frac{2}{3}\delta_{ij}{\bf w}\cdot {\bf r}\right) dV}.
\end{equation}
 Figure \ref{fig:HeModes} shows the first few pressure modes of the cylindrical cavity that have a non-zero quadrupolar tensor, along with the form of the tensor. As can be deduced from eq. \ref{eq:potential}, several modes have a zero quadrupole moment due to symmetry.

In their analysis of various antenna geometries for gravitational radiation detection, Hirakawa and co workers introduced two quantities to compare GW antennas spanning different size and symmetry groups \cite{Hirakawa1976}. These are the effective area of the antenna ($A_G$) characterizing the GW-active part of the vibrational mode, and the directivity function ($d^A$), which characterizes the directional and polarization dependence of the GW sensor. They are defined as 
\begin{equation}
A_G = \frac{2}{\mu M}\sum q_{ij}^2
\end{equation}
and 
\begin{equation}
d^A\left(\theta,\phi \right) = \frac{5}{4}\frac{\left( \sum q_{ij}e_{ij}^A\left({\bf k}\right) \right)^2 }{\sum q_{ij}^2}
\end{equation}
where $M$ is the total mass of the antenna and $ e_{ij}^A$ is the unit vector for incoming GW signal polarization $A$ ($A\in \{+, \times\}$) in arbitrary direction ${\bf k}( \theta, \phi, \psi)$. The Euler angles $( \theta, \phi, \psi)$ transform from the pulsar coordinate system to the detector co-ordinate system and are discussed in Appendix A, along with the explicit form of $ e_{ij}^A$. In sum, the angles $\theta$ and $\phi$ describe the direction of the incoming gravitational wave, and $\psi$ defines the polarization of the detector (rotation of the $x-y$ plane of the source). An important distinction between the proposed detector and other gravitational wave sensors, particularly the interferometric ones is that the orientation of the detector can be adjusted to optimize the directivity function for the astrophysical source in consideration due to its small size. This acts as another tunable parameter that can give significant enhancement in sensitivity for a particular source, as shown in fig. \ref{fig:He020Directivity} for a specific acoustic mode $(l,m,n)=(0,2,0)$. 

In terms of previously defined expressions, the mean squared signal force from a continuous gravity wave source of polarization $A$ is given by
\begin{eqnarray}
\overline{f_G^2} &=& \frac{2\pi G}{5c^3}M \mu \omega_G^2 A_G d^A \left( \theta, \phi \right) s^A,\\
&=&\frac{1}{40}M \mu \omega_G^4 A_G d^A {\overline {h_A \left( t \right)^2} }
\end{eqnarray}
where $\omega_G=2\omega_p$ is the frequency of gravity wave. Here, we have assumed a delta-function gravity wave spectrum.

\begin{widetext}
\begin{center}
\begin{figure}
\includegraphics[width=6in]{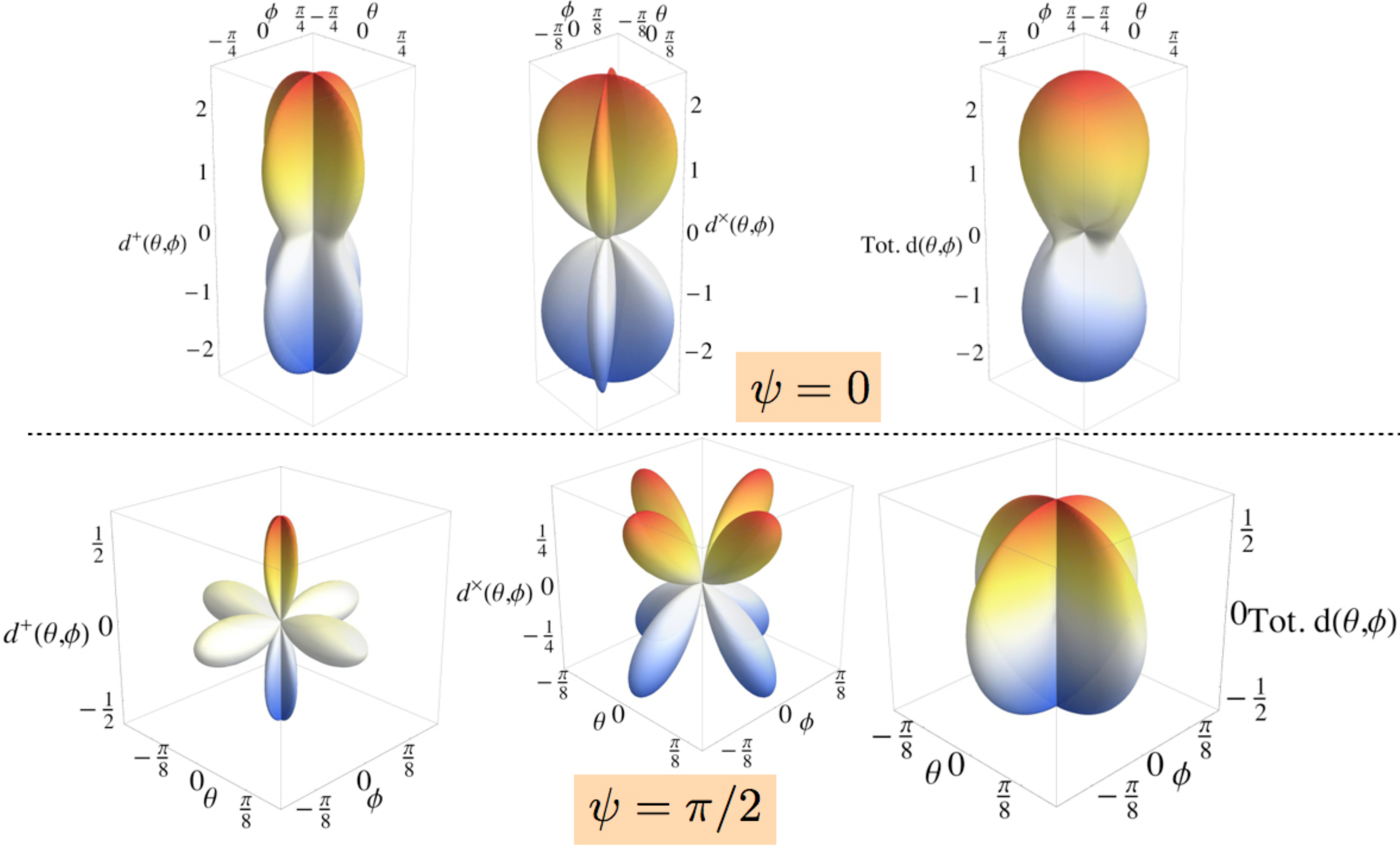}
\caption{ The directivity patterns for  acoustic mode $(l,m,n)=(0,2,0)$ of the cylindrical cavity. The $+,\times$ polarization and total directivity functions are given for two different polarizations of the detector (Euler angle $\psi=0,\pi/2$). The orientation of the detector can be adjusted to optimize the directivity function for the astrophysical source in consideration.}
\label{fig:He020Directivity}
\end{figure}
\end{center}
\end{widetext}
 As an example, we choose a cylindrical cavity of radius $a=10.8$ cm, length $L=50$ cm (from now onwards referred to as Gen1 or with subscript $(\rm He,1)$). We focus on acoustic mode $f_{(0,2,0)}=1071$ Hz, which has an effective mass $\mu = 0.625 M$, and a large GR-active area of $A_{G}=0.629 \pi a^2$ due to its quadrupolar shape, shown in figure \ref{fig:HeModes}. Another geometry considered in this work is a cylindrical cavity of the same radius,  but length $L=3$ m (from now onwards referred to as Gen2 or with subscript $\rm( He, 2)$). Since the resonance frequency of the $[020]$ mode is independent of length, it has the same frequency. However, increasing the mass gives us a larger effective mass for the same area. Figure \ref{fig:He020Directivity} shows the various directivity functions for this acoustic mode that capture the angular dependence of the sensitivity of the detector.

\section{Noise mechanisms and minimum detectable strain}
\label{Sec:Noise_Hmin}
The system we are proposing and have been exploring in the laboratory\cite{DeLorenzo2014} is a parametric transducer\cite{bocko1984phase} and essentially similar to other optomechanical systems\cite{aspelmeyer2012quantum}: the acoustic motion of the superfluid and resulting perturbation of the dielectric constant modulates the frequency of a high-$Q$ superconducting microwave resonator.  The details of the coupled acoustic and microwave system, sources of dissipation (phonon scattering, effect of isotopic impurities, radiation loss,) microwave and signal detection limits, effects of electrical dissipation, requirements on thermal stability, etc. will be the subject of another manuscript\cite{DeLorenzo2016}.  Here we take a few central results of this analysis.  

The noise sources relevant to this system are the Brownian motion of the fluid driven by thermal/dissipative forces, the additive noise of the amplifier which is used to detect the microwave field, the added noise of the stimulating microwave field (phase noise), and possible back-action forces due to fluctuations of the field inside the microwave cavity (due to phase noise and quantum noise). We will assume for the purpose of this discussion that the challenging job of seismically isolating the superfluid cell from external vibrations has been accomplished as has been done for other gravitational wave detectors. Due to the high frequency and narrow bandwidth of the astrophysical source of interest, the strain noise due to Newtonian gravity fluctuations are expected not to be relevant for this detector \cite{Adhikari2014}. The effect of vortices in superfluid helium due to earth's rotation on the $Q$-factor is unclear. However, using an annular cylinder or an equatorial mount allows for long integration times without the possibly detrimental effects due to vortices.

For a sufficiently intense microwave pump, with sufficiently low phase noise, the thermal Brownian motion of the helium will dominate the noise.  Assuming the device is pumped on the red-sideband, $\omega_{pp} = \omega_c - \omega_m$, and that the system is the side-band resolved limit, $\omega_m > \kappa_c$, the upconversion rate of microwave photons is given by: $\Gamma_{opt}=4(\Delta p_{SQL}\cdot g_0)^2 n_p /\kappa_c$, where $\omega_{pp}$, $\omega_c$, and $\omega_m$ are the pump, cavity, and acoustic mode frequency respectively, $\kappa_c=\omega_c/Q_{Nb}$ is the cavity damping rate, $\Delta p_{SQL}$ is the amplitude of the zero-point fluctuation of the pressure of the acoustic field, $n_p$ is the amplitude of the pump inside the cavity measured in quanta, and $g_0$ is the coupling between the acoustic and microwave field.  For the geometry we consider here, Gen1: $l=0.5$ m, $d=0.108$ m, $\omega_m=1071 \cdot 2 \pi $ Hz, $\omega_c= 1.6\cdot 2\pi$ GHz, and $g_0=-7.5 \cdot 10^{-11} \cdot 2 \pi $ Hz.

To achieve a readout with noise temperature of 1mK, which means that the added noise of the amplifier is equal to the thermal noise amplitude when the helium is thermalized at 1mK, requires $n_p=6\cdot 10^{9}$ microwave pump photons and a phase noise of $-145$ $db_c$/Hz.  To begin to dampen and cool the acoustic resonance with cavity backaction force, would require $n_p=10^{12}$, and a phase noise of $-145$ $db_c$/Hz. Microwave sources have been realized using whispering gallery modes of sapphire with phase noise of $-180$ $db_c$/Hz.  Together with a tunable superconducting cavity, it is possible to realize a source with sufficient low noise to broaden and cool this mode with backaction.  Furthermore, as we will detail in our future work\cite{DeLorenzo2016}, $^3$He impurities diluted into the $^4$He are expected to add acoustic loss, additional to the 3-phonon process.  To achieve $Q_{He}=10^{11}$, we estimate that an isotopic purity of $n_{3}/n_{4}=10^{-11}$ is required.  

Due to the very low dielectric constant of helium ($\epsilon_{He}=1.05$), the bare optomechanical coupling constant is small compared to typical micro-scale optomechanical systems: $g_0=\Delta p_{SQL}\cdot\partial \omega_{c}/\partial \Delta p=-7.5 \cdot 10^{-11} \cdot 2 \pi $ Hz: this is the frequency shift of the Nb cavity, $\omega_c$, due to the zero-point fluctuations of the acoustic field of the helium, $\Delta p_{SQL}$.  However, the relevant quantity is cooperativity, $C=\Gamma_{opt}/\gamma_{He}$, which compares the rate of signal photon up-conversion, $\Gamma_{opt}$, to the loss rate of acoustic quanta to the thermal bath, $\gamma_{He}=\omega_{He}/Q_{He}$. With quantum limited microwave detection (now possible with a number of new amplifiers), detection at the SQL is achieved when $C=1$, and is the onset of significant backaction effects such as optomechanical damping and cooling.  The key point is that for this system we expect to be able to realize very large $n_p$.  This is due to the very high $Q$ possible in Nb, ($Q_{Nb}\sim 10^{11}$ is now routine for accelerator cavities \cite{Allen1971,Eichhorn2014}, even when driven to very high internal fields of $10^{7}V/m$ corresponding to $n_p=10^{23}$,) and dielectric losses and resulting heating at microwave frequency in liquid helium are expected to be negligible up to very high pump powers. Assuming the dielectric loss angle in helium is less than $10^{-10}$, our estimates suggest that $n_p=10^{16}$ should be achievable before dissipative effects lead to significant heating of the helium sample at 5mK, far beyond the internal pump intensity used with micro-optomechanical systems and far above the onset of backaction effects, $C=1$ for $n_p=8 \cdot 10^{11}$.  As a result, we are optimistic that SQL limited detection and significant backaction cooling and linewidth broadening are possible. 

Since the frequency and phase of the pulsar's gravity wave signal should be known through observations of the electromagnetic signal, single quadrature back-action evading, quantum non-demolition measurement techniques could be implemented\cite{Suh2014}.  This has the advantage of avoiding the back-action forces from the cavity field fluctuations and can lower the phase noise requirements of the microwave pump.

For a damped harmonic oscillator with $\gamma_{He} = \omega_m/Q_{He}$ in equilibrium with a thermal bath at overall effective temperature $T$, the position noise spectral density is given by
\begin{eqnarray} \nonumber
 S^{\rm th}_{\xi\xi}[\omega]=\frac{k_B T}{\mu \omega_m^2} \left\{ \frac{\gamma_{He}/2}{(\omega+\omega_m)^2+\gamma_{He}^2/4 }\right. \\
 +\left.\frac{\gamma_{He}/2}{(\omega-\omega_m)^2 +\gamma_{He}^2/4}\right\}.
\end{eqnarray}
Assuming that noise at the detection frequency is dominated by the thermal noise of the acoustic mode, the force noise spectral density $S_{FF}$ is given by the relation $S_{\xi\xi}[\omega]=|\chi(\omega)|^2 S_{FF}[\omega]$, with the susceptibility $\chi(\omega) = [\mu((\omega_m^2 -\omega^2)+i\gamma_{He}\omega)]^{-1}$.

For gravitational strain, using eq. \ref{eq:EOM}, we find $S_{hh}[\omega] = 40 S_{FF}[\omega]/(\mu M \omega_G^4 d^A A_G)$ for a continuous gravity wave source at frequency $\omega_G$. Combining these, we find that for a resonant mass detector at $\omega_G = \omega_m$,
\begin{equation}
S_{hh}[\omega] = \frac{80 k_B T}{M d^A A_G Q_{He} \omega_m^3}.
\end{equation}

The strain sensitivity of our detector is simply $\sqrt{S_{hh}[\omega]}$, and the minimum noise is $\sqrt{S_{hh}[\omega]/\tau_{\rm int}}$ after an integration time $\tau_{\rm int}$. 
The minimum detectable strain field with $2\sigma$ certainty is therefore given by \cite{New1995}
\begin{equation}
h_{min}\approx 2\sqrt{\frac{S_{hh}[\omega]}{\tau_{\rm int}}}=\sqrt{\frac{320 k_B T }{M\omega_{G}^3 A_G d^A Q_m}\frac{1}{\tau_{\rm int}}}.
\label{eq:Hedet_hmin}
\end{equation}
The $2\sigma$ uncertainty limit is used to be consistent with previously reported limits on $h_{min}$ set by LIGO \cite{Aasi2014}.

As an example, both cylindrical cavities considered in section \ref{Sec:HeliumDet} have acoustic mode $f_{(0,2,0)}\sim1071$ Hz. This mode of the detector can easily be tuned (by under $\pm$15 Hz) to be in resonance with pulsars J0034-0534, J1301+0833, and J1843-1113.  Similarly, another acoustic mode ($f_{[2,0,1]}= 1425$ Hz) is found to have resonant frequencies in the vicinity ($<$8 Hz) of the frequency of gravitational waves from pulsar J1748-2446ad. Taking into account the different quadrupole tensors, effective mass and directivity functions for the different acoustic modes, Table 2 lists the minimum detectible strain for several pulsars for cylindrical detector Gen1 after 250 days of integration time (same time as the current LIGO+VIRGO estimates in Ref. \cite{Aasi2014}). Here we have assumed an acoustic $Q$-factor of $ 10^{11}$ and thermal $T_{th} = 5$ mK 
for both geometries. Since the detector is small enough to be rotated or moved geographically to optimize signal from a particular pulsar, we have assumed $\psi=0$ and $(\theta,\phi)$ that maximizes the directivity.

\begin{widetext}
\begin{center}
\begin{table}[ht]
\label{tab:MSPtable}
\begin{tabular}{|c |c| c| c| c|c|c|}
\hline
Pulsar & $\omega_p/2\pi$ & $f_{GW} $ (Hz) &$h_{sd}$ & $h_0^{95\%}$- LIGO & $h_{\rm He,1}^{95\%}$[l,m,n] &$h_{\rm He,2}^{95\%}$[l,m,n] \\
\hline\hline

J0034-0534& 532.71 & 1065.43  & $2.7 \times 10^{-27}$  &$1.8 \times 10^{-25}$ & $1.1\times10^{-26}$[020]& $3.6 \times 10^{-27}$[020]\\
\hline

J1301+0833 & 542.38 & 1084.76 & $2.8 \times 10^{-27} $ & $1.1 \times 10^{-25} $ & $1.0\times10^{-26}$[020]& $3.5 \times10^{-27}$[020]\\
\hline

J1843-1113 & 541.81 & 1083.62 & $1.1 \times 10^{-27} $ & $1.1 \times 10^{-25} $ & $1.0 \times 10^{-26}$[020] & $3.5 \times 10^{-27}$ [020]\\
\hline

J1748-2446ad & 716.36 & 1432.7 & $6.7 \times 10^{-28} $ & $1.8\times 10^{-25}$ & $1.3\times10^{-26}$[201]& no coupling\\
\hline
\end{tabular}
\caption{Table of millisecond Pulsars of interest for helium detector Geo1 and Geo2. Here, $\omega_p$ is the oulsar rotational frequency, $ f_{GW} = \omega_p/\pi$ is the frequency of gravitational waves, as given in Table 1. These values are compared to the strain limit set by recent continuous GW survey by interferometric detectors \cite{Aasi2014}, along with the strain limit for the helium resonant detectors Gen1 with an integration time of 250 days and Gen 2 with one year integration time, as shown in eq. \ref{eq:Hedet_hmin}. Here, $\psi = 0$, $Q$-factor is $10^{11}$, and the acoustic mode is given in square brackets.}
\end{table}
\end{center}
\end{widetext}

In order to compare the sensitivity of our proposed detector with other GW sensors, we pick a specific expected astrophysical source: gravity waves from pulsar J1301+0833, with $\omega_G = 2 \pi \times 1084.76$ Hz. Gen1 (mode [020]) gives us sensitivity of $h_{min} = 3.4 \times 10^{-23}/\sqrt{\rm Hz}$, which is significantly below the sensitivity of LIGO, and comparable (within a factor of 2) to current sensitivity of advanced LIGO. Such a detector can surpass the LIGO +VIRGO estimate on minimum strain $ h_0^{95\%} = 1.1 \times 10^{-25}$ in under a week of integration time (under a month if $Q$-factor is $10^{10}$ instead). Increasing the mass by a factor of 6 (by choosing Gen2), while assuming the same $Q$-factor and noise characteristics, we can get sensitivity of $1.4 \times 10^{-23}/\sqrt{\rm Hz}$, which is below the strain sensitivity of advanced LIGO for this frequency range. Figure \ref{fig:hmin} shows the minimum detectable strain as a function of integration time for various $Q$ factors for  two resonant detectors operating at the same sensitivity. Figure \ref{fig:hmin} also shows the sensitivity estimates for three interferometeric detectors operating at LIGO-S6 sensitivity, and at advanced LIGO design sensitivity, as used in ref. \cite{Aasi2014}.  

As figure \ref{fig:hmin} and Table 2 demonstrate, Gen2 can come within a factor of 2 of the spin-down limit for pulsar J1301+0833 (and several other pulsars) in a year of integration time. Considering the conjecture that the primary spin-down mechanism for MSPs is the emission of gravitational radiation, our detector seems a promising candidate for searches of continuous gravitational waves from this and similar other pulsars.  
\begin{figure}[ht]
\begin{center}
\includegraphics[width=3.4in]{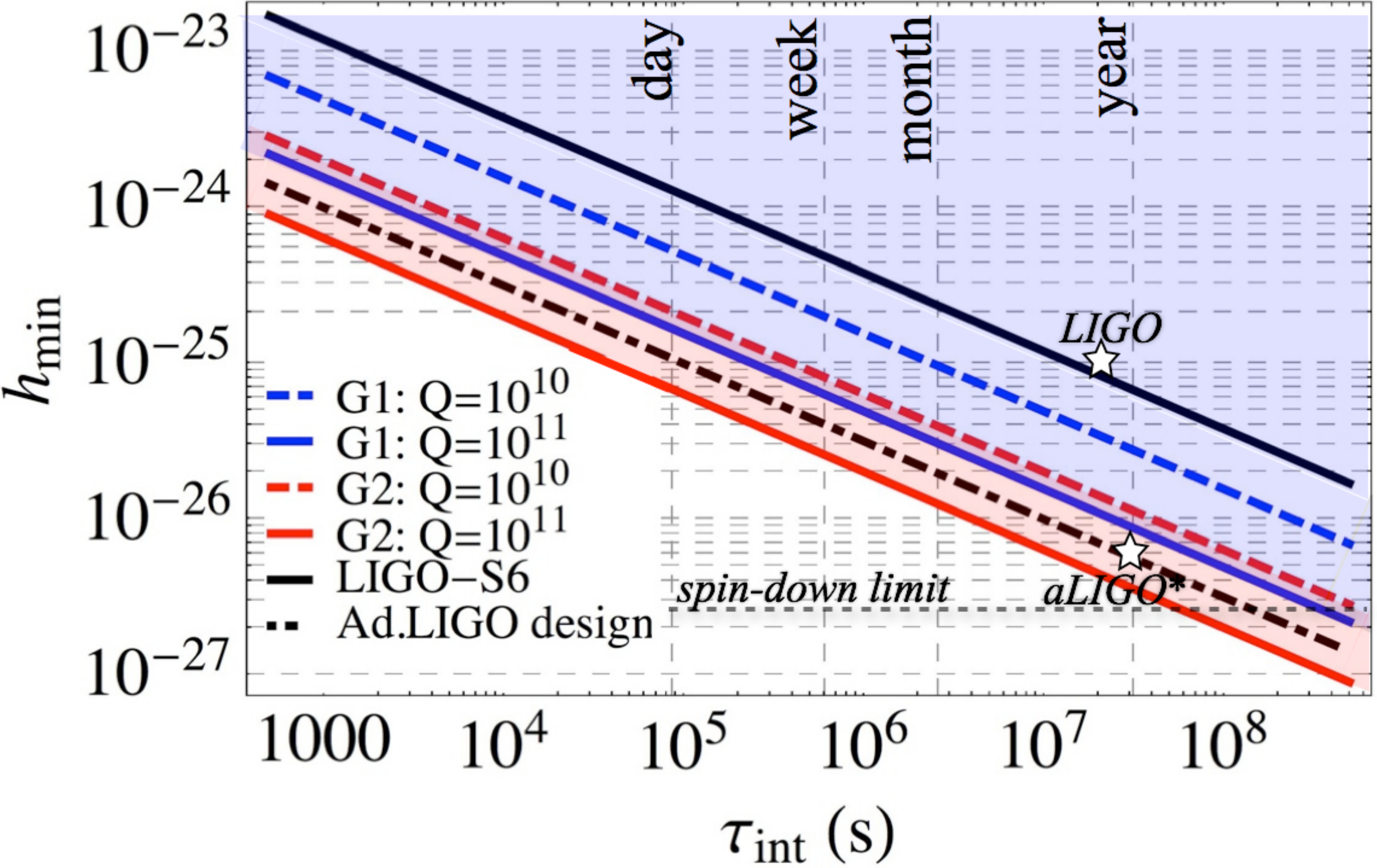}
\caption{Strain sensitivity versus measurement time for two helium detectors with same sensitivity operating simultaneously, assuming a bath temperature of 5mK, $(l,m,n)=(0,2,0)$, for Geometry 1(blue) and 2(red). We also show the limits set by three interferometric detectors operating at LIGO-S6 sensitivity (solid black), and the design sensitivity of advanced LIGO (dashed black). The stars shows the current limit on minimum strain set by LIGO, and the projected limit by Advanced LIGO. As seen in the figure, the current limit can be surpassed within a few days of integration time for Gen1, and under a day for Gen2. Also shown is the spin-down limit for pulsar J1301+0833.}
\label{fig:hmin}
\end{center}
\end{figure}

We would like to note that several noise suppression mechanisms (such as squeezed light injection) currently used in LIGO can also be employed here. More importantly, there are ways to squeeze the mechanical motion of the detector \cite{Wollman2015,Pirkkalainen2015,Lecocq2015}. This can significantly relax the size, $Q$-factor and microwave noise requirements, increasing the sensitivity of our proposed detector significantly. For example, exploring methods to squeeze mechanical motion by changing the speed of sound periodically, and exploring other effects arising from parametric coupling between the helium acoustic modes and the microwave resonator container is a straight forward extension of the current setup, since the helium is already being pressurized and parametrically coupled to microwaves for resonant force detection. A detailed analysis of implementing these protocols for improved gravity wave sensing will be the subject of future research.

\section{Comparison with other detectors}
\label{Sec:Compare}

The basic principle of the superfluid helium detector is analogous to that of other resonant mass sensors, such as Weber bars. The use of resonant mass GW detectors has a 50 year history, dating back to early experiments by Weber \cite{Weber1960}. There have been several proposals of using resonant mass detectors to search for GW from pulsars, for example Ref. \cite{New1995}, and a few continuous GW searches targeting specific pulsars, the most notable one being the Tokyo group experiment looking for signal from Crab Pulsar\cite{Tsubono1991}. Here we highlight several key differences in the implementation  using superfluid helium. 

\begin{itemize}
{\item {\underline{Mass:}} We discuss a kg-scale sample of helium which is $10^3$ times smaller than the typical resonant bar detectors. The low mass limits the utility of the helium detector to CW sources, where as the massive detectors are useful for burst sources. Nonetheless, there is high sensitivity for CW sources and the low mass makes a helium detector economical and small scale. One could deploy a few such detectors to seek coincidence and further improve sensitivity. }

{\item {\underline{$T/Q_{He}$ – temperature and quality factor:}} It is possible to cool an isolated sample of helium to temperatures less than 10mK and we are anticipating very low loss. For instance, helium at 25mK with $Q_{He} = 10^9$ has a ratio $T/Q_{He}\ 10^3$ times smaller than the best value found in the literature, and potentially $10^6$ times smaller at lower temperature \cite{DeLorenzo2016}.  }

{\item  {\underline{Optomechanical damping:}} It appears possible to substantially increase the acoustic resonance linewidth without decreasing the force sensitivity by parametrically coupling to microwaves \cite{DeLorenzo2014}. While parametric transducers are also used in other resonant mass detectors \cite{Tobar1993, Tobar2000}, the particular geometry and mechanism used in helium detector is expected to have lower noise characteristics \cite{DeLorenzo2016}.}

{\item  {\underline{Frequency tunability:}} It is possible to change the speed of sound in helium by 50\% by pressurization. This allows the apparatus to be frequency agile; thus searching several pulsars with the same detector. It also allows for long term tracking the same pulsar in the presence of deleterious frequency shifts. For example, the estimated Doppler shift of the GW signal from Crab pulsar is $\sim 30$ mHz/year due to earth's motion. Our detector can be tuned to track this shift, allowing for months of integration time. By resonantly tracking the pulsar we also reduce SNR, and thereby the detection threshold. }

\end{itemize}

A standard figure of merit used in literature to compare various bar detectors of different materials is  $\eta= Q \rho c_{s}^3$ \cite{Ju2000}. Typical values of $\eta$ range from $10^{21}-10^{24}$ kg s$^{-3}$. According to this metric, helium may seem like a poor choice for a bar detector, ($\eta\sim Q_{He}\times10^9$ kg s$^{-3}$). This figure of merit is made of the material specific parameters in the minimum detectable strain, as given in eq. \ref{eq:Hedet_hmin}. However, adding the temperature dependence, and the significantly large $Q$-factors make the helium sensor comparable to the resonant bar detector. In addition, due to it's smaller size, temperature stability, seismic and acoustic isolation are much easier to maintain. 

Unlike interferometric detectors like LIGO conducting a broadband search for gravitational waves, the helium detector is narrowband, and works best for detection of continuous waves such as pulsars. Nevertheless, as highlighted in fig. \ref{fig:hmin}, around 1kHz the setup described above has strain sensitivity within a factor of 4 (Gen 1), or in principle even surpassing the sensitivity of advanced LIGO by considering a larger volume of superfluid helium (Gen 2). This allows us to surpass the limits from previous CW searches of VIRGO+LIGO experiments ($h_{min}\sim 10^{-25}$) within a week, or less depending on the detector size and $Q$-factor.

There are several ongoing and proposed detectors for gravitational waves, for example space-based interferometric detector eLISA \cite{Vitale2014, Armano2016}, atom interferometry based detector AGIS-LEO \cite{Hogan2011}, 
and Pulsar Timing Arrays \cite{Hobbs2010}. These detectors operate at different frequency ranges, typically much lower than the ones considered here. The astrophysical sources of interest are therefore different from those of the helium detector. 

Finally, an important advantage of considering superfluid helium as a resonant GW sensor is that by designing different geometries and exploring different types of resonances, one could build detectors for a range of astrophysical sources. For example, by considering smaller containers or Helmholtz resonances in micro or nano-fluidic channels \cite{Rojas2015}, it may be possible to build a resonant detectors for high frequency sources of gravity waves as explored in other devices \cite{Arvanitaki2013,Goryachev2014}. Alternatively, larger containers or low-frequency Helmholtz resonances may be used to detect continuous GWs from young pulsars or binary systems. Since the technology required for the proposed superfluid helium gravity detector is space-friendly, it may be possible to design low frequency detectors for space missions if seismic noise becomes a deterrent.

\section{Conclusions and Outlook}
\label{Sec:Conclusion}

As discussed in Section \ref{Sec:Noise_Hmin}, there are several stringent requirements for low-noise operation of our proposed helium detector: isotopically pure sample, sub-10mK cryogenic environment, very low phase-noise microwave source, and isolation from environmental vibrations. Furthermore, due to the low density and speed of sound, a reasonable size ($\sim1$m) bar detector made of helium can only be used for detection of continuous gravity waves.

Despite these extreme requirements, using  superfluid $^4$He does have several advantages. The low intrinsic dissipation and dielectric loss and wide acoustic tunability are direct manifestations of the inherent quantum nature of the acoustic medium. Furthermore, due to the mismatch between the speed of sound in helium and niobium, there is an inherent acoustic isolation from the container. Since the container itself is in a macroscopic quantum state (superconductor), it further contributes to the extremely low-noise, high sensitivity nature of the proposed device by making an extremely high $Q$ microwave resonator with very high power-handling.   

Several ideas for future work are outlined in the manuscript at various places. They include investigating more complex geometries for stronger coupling to gravitational strain, or investigating other high-$Q$ acoustic resonances (Helmholtz resonances) in helium to detect other sources of continuous gravity waves. Also, many ideas from quantum optics and quantum measurement theory can be implemented in this system to increase bandwidth or sensitivity. For example, by periodically modulating the acoustic resonance frequency, it will be possible to upconvert out of resonance signals into helium resonance signals, thereby increasing the frequency tunability of our detector. Several techniques from quantum measurements can be applied to our proposed transduction scheme to avoid measurement backaction or to squeeze acoustic noise, thereby increasing the sensitivity further.  

Even without these techniques, the extreme displacement sensitivity ($\sim 10^{-23}/\sqrt{Hz}$) of this meter-scale device corresponds to a measurement of the width of milky way to cm-scale precision! This is again made possible by combining two macroscopic quantum states in the measurement scheme (a superfluid coupled to a superconductor). The resulting hybrid quantum sensor is an extremely low noise detector at low temperatures due to the robustness of the quantum state involved. As these experiments develop, proposing a more broadly functioning gravity wave detector may be possible, as well as the detection of other extremely small laboratory forces.

\begin{acknowledgments}
We would like to acknowledge helpful conversations with Rana Adhikari, Yanbei Chen, Dan Lathrop, Pierre Meystre, David Blair and Nergis Mavalvala. We acknowledge funding provided by the Institute for Quantum Information and Matter, an NSF Physics Frontiers Center (NSF IQIM-1125565) with support of the Gordon and Betty Moore Foundation (GBMF-1250) NSF DMR-1052647, the NSF ITAMP grant, and DARPA-QUANTUM HR0011-10-1-0066. 
\end{acknowledgments}

\appendix

\section{Brief introduction to Gravitational waves}

Gravitational waves are solutions to the linearized Einstein equations, where the perturbed metric can be written as $g_{\mu \nu} = \eta_{\mu \nu} + h_{\mu \nu}$. Here, $\eta_{\mu \nu}=\textrm{diag}[-1,1,1,1]$ is the Minkowski metric (which is a good approximation for our solar system) and $|h_{\mu \nu}| \ll 1$ is a small perturbation of the metric.  In free space, Einstein's equations of motion, which describe the dynamics of space-time, reduce to $R_{\mu \nu} = 0$, where $R_{\mu \nu}$ is the Ricci-tensor constructed from the metric. Since only the weak-field limit is considered, terms that are of higher order in $h_{\mu \nu}$ can be neglected. In addition, general relativity has an inherent gauge freedom related to the choice of coordinates. In the Lorentz-gauge the equations of motion reduce to a wave equation as in electromagnetism:  
\begin{equation}
\label{eq:wave}
R_{\mu \nu} = \Box h_{\mu \nu} = \left( -   \partial_t^2 + c^2 \nabla^2\right) h_{\mu \nu} = 0 \, .
\end{equation}

This is the wave equation for gravitational waves, which are small perturbations of flat space-time that propagate at the speed of light. A general plane-wave solution has the form $h_{\mu \nu}(\vec{x}, t) = A_{\mu \nu} \cos(\omega t - \vec{k} \cdot \vec{x} +\varphi)$, with the dispersion relation $\omega = c|\vec{k}|$. 
Choosing the specific transverse-traceless gauge, and a coordinate system in which the wave propagates only in the z-direction, the only non-vanishing components of the gravitational wave tensor are the spatial components
\begin{equation}
h_{i j} = h_+\left(t-\frac{z}{c}\right) {\bf e}^{+}_{ij}\left(\hat z\right) + h_{\times} \left(t-\frac{z}{c}\right) {\bf e}^{\times}_{ij}\left(\hat z\right)
\end{equation}
%
where $h_+$ and $h_{\times}$ are the two polarization components with the polarization tensors given by
\begin{equation}
\bf{e}^+\left(\hat z\right)=
\left(
\begin{array}{ccc}
1 & 0 & 0 \\
0 & -1 & 0 \\
0 & 0 & 0 \\

\end{array}
\right)
\label{eq:eplus}
\end{equation}
and
\begin{equation}
\bf{e}^x\left(\hat z\right)=
\left(
\begin{array}{ccc}
0 & 1 & 0 \\
1 & 0 & 0 \\
0 & 0 & 0 \\

\end{array}
\right)
\label{eq:ecross}
\end{equation}

Gravitational waves carry energy and have observable effects on matter. For test particles at a distance much shorter than the wavelength of the gravitational wave, the wave induces an effective time-dependent tidal force. To see this, it is convenient to use gauge-invariant quantities, such as the Riemann tensor which is invariant to linear order. Its only non-vanishing component is $R_{\mu 0 \nu0} = -\frac{1}{2} \ddot{h}_{\mu \nu}$, where the dot denotes differentiation with respect to coordinate time $t$. The Riemann tensor captures how neighboring geodesics (i.e. world lines of free particles) change with respect to each other: the vector $x^{\mu}$ that connects two geodesics follows the geodesic deviation equation $\ddot{x}^{\mu} = R^{\mu}_{0 \nu 0} x^{\nu} = -\frac{1}{2} \ddot{h}_{\mu \nu} x^{\nu}$. This equation holds for geodesics that are close to each other as compared to the wave length $\lambda$ of the gravitational wave, i.e. $x << \lambda$. From this equation follows the equation of motion for the distance between two neighboring test particles:
\begin{equation}
\begin{split}
\ddot{x} & = \frac{1}{2} \left( \ddot{h}_+ x + \ddot{h}_{\times} y \right) , \\
\ddot{y}& = \frac{1}{2} \left(  \ddot{h}_{\times} x - \ddot{h}_+ y \right) .
\end{split}
\end{equation}\label{eq:deviationsDiff}
The equations of motion are equivalent to the presence of an effective tidal force $F_i = \ddot{h}_{i j} x^j/2$ that acts on the particles. The corresponding effective force is conservative and can therefore be represented by force lines, shown in fig. \ref{fig_potential} for a purely plus-polarized wave. For a general polarization, the force line diagram is rotated counter-clockwise by the angle $\Psi$ where $\tan(2 \Psi) = \ddot{h}_{\times} / \ddot{h}_+$.

For only a plus-polarized wave ($h_{\times}=0$), the solution to lowest order in $h$ is
\begin{equation}\label{eq:deviations}
\begin{split}
x(t) & = x(0) \left(1 + \frac{h_+(t)}{2} \right)  \\
y(t) & = y(0) \left(1 - \frac{h_+(t)}{2} \right).
\end{split}
\end{equation}

The distances between nearby points oscillate in the $x-$ and $y-$directions, i.e. perpendicular to the gravitational wave. A cross-polarized wave has the same effect but with the x-y-plane rotated by $\pi/4$ (see also fig. \ref{fig_wave}).
\begin{figure}[ht!]
\centering
\includegraphics[width=0.5\columnwidth]{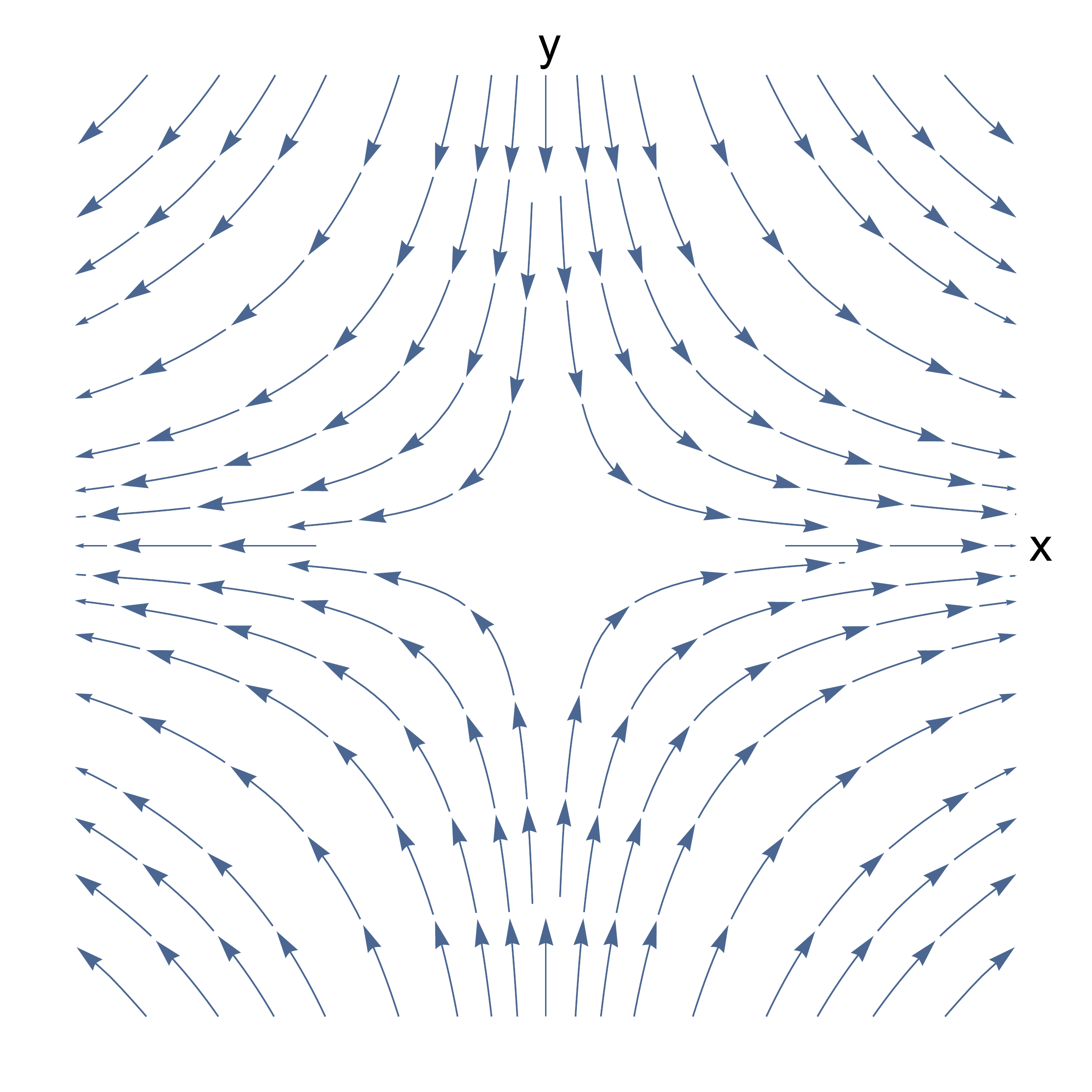}
\caption{\small Force lines for the effective tidal force produced by a plus-polarized gravitational wave \eqref{eq:deviationsDiff}. The force acts perpendicular to the direction of propagation of the wave. The same force lines, rotated counter clock-wise by $\Psi = \pi/4$, represent the effect of a cross-polarized wave. }
\label{fig_potential}
\end{figure}
\begin{figure}[hb]
\centering
\includegraphics[width=0.85\columnwidth]{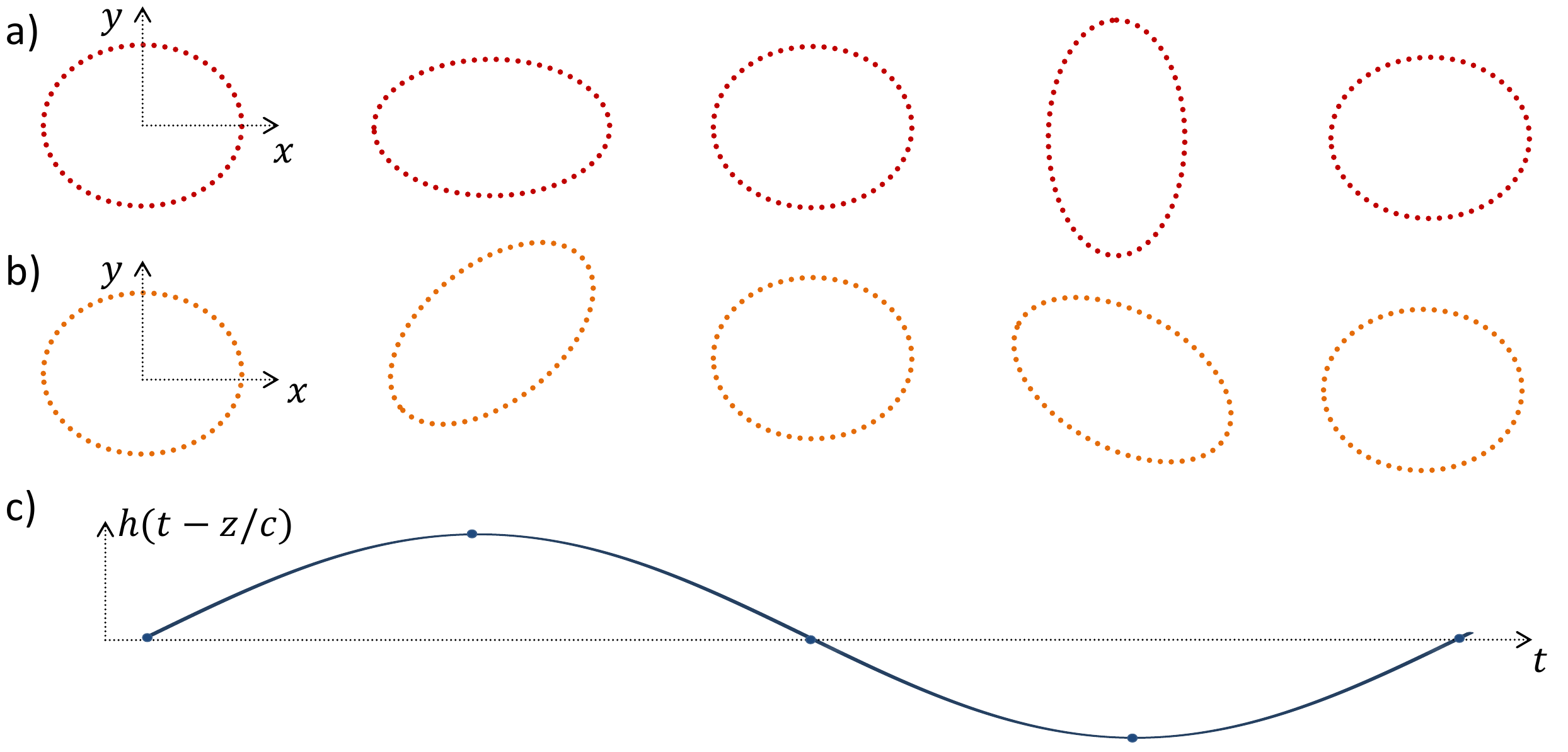}
\caption{\small Dynamics of mass distribution in time, as the gravitational wave passes. Row (a) shows a ring of test particles for a passing plus-polarized wave, while row (b) shows the effect of a cross-polarized wave.}
\label{fig_wave}
\end{figure}
%

\begin{figure}[ht]
\begin{center}
\includegraphics[width=0.6\columnwidth]{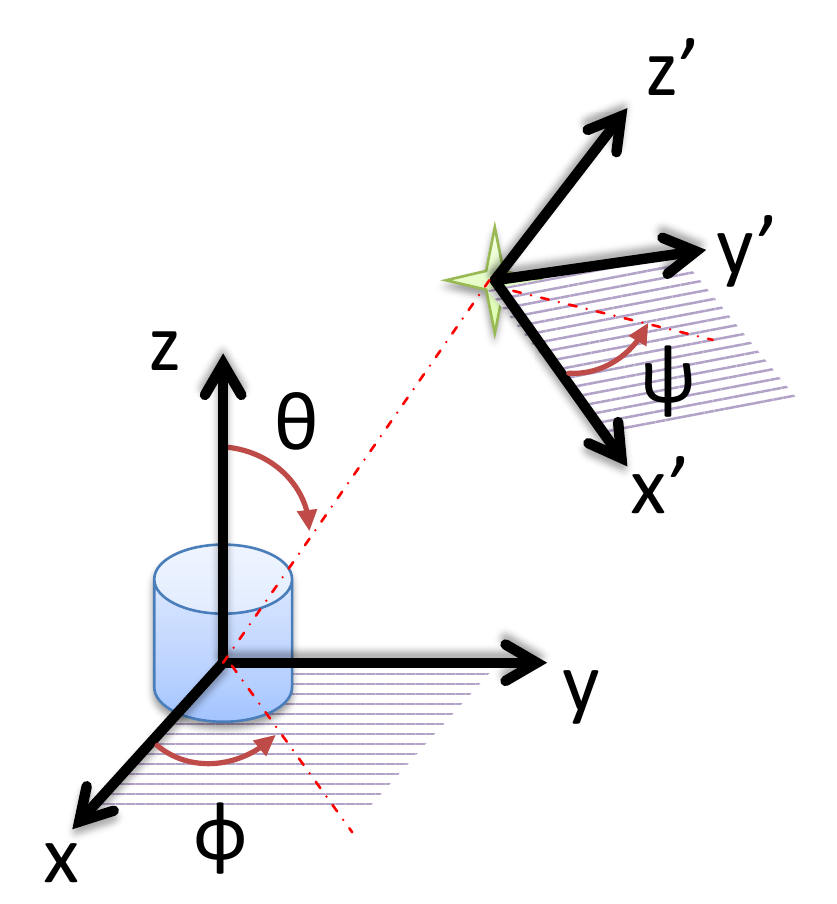}
\caption{ The co-ordinate transformation angles from the detector frame (symbolized by the cylinder) to the source frame (symbolized by the star). The angle $\Psi$ in the x'-y'-plane defines the polarization of the gravitational wave. 
}
\label{fig:Euler}
\end{center}
\end{figure}

The detector co-ordinate axis ($x,y,z$) is not necessarily aligned with the gravitational wavefront emitted from the source ($x',y',z'$). To account for the angular dependence, the strain at the detector can be written as
\begin{equation}
h(t)=F_+(\theta,\phi,\psi)h_+(t)+F_\times(\theta,\phi,\psi)h_\times(t)
\end{equation}
where $(\theta,\phi,\psi)$ are the Euler angles that convert from the pulsar co-ordinate system to the the detector plane, as shown in fig. \ref{fig:Euler}, and $F_{+/\times}(\theta,\phi,\psi)$ are known as the detector pattern functions \cite{Bonazzola1996}. While the angles $\theta$ and $\phi$ describe the direction of the incoming gravitational wave ($\phi$ being rotation of the old $x-y$ plane along the z-axis, and $\theta$ being the angle between the source and detector z-axis), $\psi$ defines the polarization of the detector (rotation of the $x-y$ plane along source line of sight) \cite{Christensen1992}, as shown is fig. \ref{fig:Euler}. Compared to large ground-based sensors where the angle $\psi$ is fixed, it can be used as a parameter for the helium detector, to be optimized for the particular pulsar in consideration.

$F_{+/\times}(\theta,\phi,\psi)$ is defined as
\begin{equation}
F_A(\theta,\phi,\psi) = q_{ij}\hat{e}_A^{ij}(\theta,\phi,\psi),
\end{equation} 
where ${\bf q}$ is the dynamic mass quadrupole tensor of the detector that we will visit later, and $\hat{e}_A^{ij}(\theta,\phi,\psi)$ (with $A\in \{+,\times\}$) are the unit vectors for the two polarizations of the gravitational wave given in eqs. \ref{eq:eplus}, \ref{eq:ecross}  in the rotated basis, $\hat{e}_A(\theta,\phi,\psi)=R^{-1}_{XYZ}\hat{e}_AR_{XYZ}$, where the rotation matrix is given by

\begin{widetext}
$ R_{XYZ}(\theta,\phi,\psi)=
\begin{bmatrix}
\cos{\theta} \cos{\phi}& \cos{\theta}\sin{\phi} & -\sin{\theta}\\
-\cos{\psi}\sin{\phi}+\cos{\phi}\sin{\theta}\sin{\psi} &\cos{\phi}\cos{\psi}+\sin{\phi}\sin{\theta}\sin{\psi} & \cos{\theta}\sin{\psi}\\
\cos{\phi}\cos{\psi}\sin{\theta} +\sin{\phi}\sin{\psi}& \cos{\psi}\sin{\theta}\sin{\phi}-\cos{\phi}\sin{\psi} &\cos{\theta}\cos{\psi}
\end{bmatrix} $ .
\end{widetext}

\section{Gravitational wave perturbation from an ellipsoidal pulsar}
As mentioned in the previous section, the generation of gravitational waves can be studied by considering the linearized Einstein equations in the presence of matter. In this section, we focus on the mechanism for generation of CW gravity waves from pulsars.

Let us assume a non-spherical pulsar, rotating about the $z$-axis. In order to emit gravitational waves, the star needs to have some asymmetry along the rotational axis (i.e. $r_x\neq r_y$). Assuming an ellipsoidal star with the axes coinciding with the principal axes of the solid of revolution, the extent of the ellipsoidal star can be described by equation
\begin{equation}
\frac{x^2}{a^2}+\frac{y^2}{b^2}+\frac{z^2}{c^2}=1,
\end{equation}
where $a,b,c$ are the semi-axes along $x,y,z$ directions respectively. We also assume a constant mass density, $\rho$. The quadrupolar mass tensor is given by \cite{CarrollBook2004}
\begin{equation}
Q_{ij}:=\rho\int_{\rm body} x_i x_j dV.
\end{equation} 
Due to the (chosen) co-ordinate system being along the principal axes, the quadrupolar tensor is given by $Q=(1/5)M_p{\rm diag}[a^2,b^2,c^2]$, where $M_p$ is the mass of the pulsar. Assuming at time t=0, $Q(t=0)={\rm diag}[Q_1,Q_2,Q_3]$, and given the star rotates about the $z$-axis with angular frequency $\omega_p$, 
\begin{equation}
Q(t)=R_z(\omega t)I(0)R^{-1}_z(\omega t)
\end{equation}
We now define two parameters, $Q=Q_1+Q_2$, and ellipticity $\epsilon=(Q_1-Q_2)/I_{zz}$, where $I_{zz}=(1/5)M(a^2+b^2)$ is the moment of inertia about the $z$-axis \cite{CreightonBook2011}. The quadrupolar tensor can now be written as
\begin{equation}
\label{eq:Qpulsar}
Q=\begin{bmatrix}
\frac{1}{2}Q+\frac{1}{2}\epsilon I_{zz}\cos{2 \omega_p t} & -\frac{1}{2}\epsilon I_{zz}\sin{2 \omega_p t} & 0\\
-\frac{1}{2}\epsilon I_{zz}\sin{2 \omega_p t}  & \frac{1}{2}Q -\frac{1}{2}\epsilon I_{zz}\cos{2 \omega_p t}  & 0\\
0 & 0 & Q_{3}
\end{bmatrix}.
\end{equation}
In the far-field limit, where size of the star (or $GM/c^2$) $\ll$ wavelength of the gravity wave ($c/\omega_p$) $\ll$ distance to detector ($d$), the gravitational wave perturbation becomes
\begin{equation}
h_{ij}(t,x)=\frac{2G}{c^4 d} \ddot{Q}(t_r),
\end{equation}
where $h$ is the gravitational perturbation tensor in transverse-traceless gauge, and $t_r=(t-d/c)$ is the retarded time, given the detector is distance $d$ away from the source. Since time retardation gives an extra overall phase here, we ignore it for our purposes. Thus,
\begin{equation}
h=\frac{2G}{c^4 d} 2\epsilon I_{zz}\omega_p^2 \begin{bmatrix}
-\cos{2 \omega_p t} & \sin{2 \omega_p t} & 0\\
\sin{2 \omega_p t}  & \cos{2 \omega_p t}  & 0\\
0 & 0 & 0
\end{bmatrix},
\end{equation}
with the two polarization components being
\begin{eqnarray}
\label{eq:hmax}
h_+&=&-\frac{4G}{c^4 d} \epsilon I_{zz}\omega_p^2\cos{2\omega_p t}\\
h_\times&=&\frac{4G}{c^4 d} \epsilon I_{zz}\omega_p^2\sin{2\omega_p t}
\end{eqnarray}

We now estimate the perturbation in the metric due to a gravitational wave from a pulsar, $h$. There are two unknowns here, $I_{zz}$ and $\epsilon$. These parameters depend on the composition of the neutron star. Even though the equation of state (relation between density and pressure) of neutron star is unknown, certain properties of a neutron stars are remarkably well understood, and agree with astronomical observations \cite{LyneBook2006}. The main bulk of the star is assumed to be a neutron fluid, with a solid crystalline crust made of heavy nuclei. A neutron star can thus be modeled as a giant nucleus, akin to a degenerate fermi gas of neutrons. Since Fermi energy is very high, temperature variations have little effect on the properties of the star \cite{LyneBook2006}. Thus, the equation of state  determining the size and radius of the star are fairly well constrained. Most models show that the radius lies between 10.5 and 11.2 km and mass ranges between 0.5$M_{\odot}$ and 3$M_{\odot}$, with all measured values close to 1.35$M_{\odot}$ (from Keplerian analysis of pulsars in binary systems- ~5\% of all observed pulsars) \cite{LyneBook2006}. Putting in these values, the moment of inertia amounts to $10^{38}$ kg-m$^2$, surprisingly close to the moment of inertia of earth! While there may be uncertainties here, the changes would amount to less than an order of magnitude in the estimate. Thus $I=10^{38}$ kg-m$^2$ is the estimate used in previous GW searches, c.f. \cite{Aasi2014}. For pulsars in binary systems of known mass, one could use a more accurate empirical expression given in Ref. \cite{Bejger2002},
\begin{eqnarray}\nonumber
I_{zz}=4.42\times10^{37} && {\rm kg\ m^2}\left(\frac{M_p}{M_{\odot}}\right)\left(\frac{R_p}{{10 \rm km}}\right)^2 \\
&&\times\left[1+5\left(\frac{M_p}{M_{\odot}}\right)\left(\frac{{\rm km}}{R_p}\right)\right], \end{eqnarray}
where $R_p$ is the average radius of the pulsar of interest. 

The biggest uncertainty in estimating $h$ therefore comes from $\epsilon$, the ellipticity parameter that characterizes the mass asymmetry of the pulsar. We now present estimates on $\epsilon$ from two different mechanisms: the spin-down energy conservation ($\epsilon_{sd}$) and elastic strain on the NS crust ($\epsilon_{n}$) and their corresponding GW strain limits.
 
{\bf Spin-down Limit: } Since the pulsar is spinning down, its rotational frequency is changing at some observable rate $\dot\omega_p$. This amounts to a torque of $I_{zz}\dot\omega_p$. If we assume that all of this spin-down is due to gravitational radiation, $I_{zz}\dot\omega_p=dL_z/dt$, where $L_z$ is the angular momentum of the body along $z$ axis. Recalling that change in energy, $\Delta E= \omega_p \Delta L_z$, we arrive at $dL_z/dt=(1/\omega_p)dE/dt$. One can derive a more rigorous expression for change in angular momentum due to gravitational radiation from first principles. But for the special case of rotation around a principal axis, the expression gets simplified to the one above \cite{CreightonBook2011}.

The emitted power of gravitational radiation is given by \cite{CarrollBook2004, CreightonBook2011}
\begin{equation}
\label{eq:GWpower}
\frac{dE}{dt}=\frac{G}{5c^5}\langle\dddot{Q}_{ij}\dddot{Q}^{ij} \rangle=\frac{32G }{5c^5}\epsilon^2 I_{zz}^2 \omega_p^6,
\end{equation}
giving us
\begin{equation}
I_{zz}\dot\omega_p=\frac{32G }{5c^5}\epsilon^2 I_{zz}^2 \omega_p^5.
\end{equation}
Substituting these values allows us to solve for the ellipticity parameter,
\begin{equation}
\epsilon_{sd}=\left(\frac{5c^5\dot\omega_p}{32GI_{zz}\omega_p^5}\right)^{1/2}.
\end{equation}  
This in turn is used to compute the upper-limit estimate for gravitation perturbation strain in terms of  constants and observational data
\begin{equation} 
\label{eq:hspindown}
h_{sd}=-\frac{4G}{c^4 d} \epsilon I_{zz}\omega^2=\sqrt{\frac{5GI_{zz}\dot\omega}{2c^3d^2\omega}}
\end{equation}

The spin-down strain estimate is a significant over-estimate of the strength of gravitational waves, particularly from young pulsars. As an example, at the surface of the Crab pulsar the ratio of gravitational to magnetic forces on an electron,
\begin{equation}
\frac{GMm}{r^2}/\frac{e\omega r B}{c}\approx 10^{-12},
\end{equation}
suggesting that electromagnetic forces have a crucial role to play in most observable properties of the star. Thus, for young pulsars it is reasonable to assume that dipolar electromagnetic radiation is the dominant way to lose angular momentum, and not quadrupolar gravitational radiation, as is eluded to by braking index measurements \cite{Manchester2005, Palomba2000}. This means that spin-down strain estimate is a significant over-estimate of the strength of gravitational waves from these pulsars, a fact already confirmed by the recent GW detector data \cite{Aasi2014}. 

A second class of pulsars, known as millisecond pulsars (or MSPs) are much longer-lived, slowly decaying, even speeding up at times. These are pulsars in binary systems with the transfer of mass (and angular momentum) from the companion star, leading to X-ray emission and speeding up of the rotations. The discovery of new MSPs has accelerated since the operation of gamma-ray detectors like Fermi-LAT. For example, 5 new MSPs were detected  by radio searches of unidentified Fermi LAT gamma-ray sources in 2012 (two of these are included in Table I) \cite{Kerr2012}.

MSPs are remarkably stable ($\dot{\omega}_p/2\pi<10^{-14}$) and were once considered strong candidates for long-term time-standard. There has only been one observed random glitch in the thousands of years of accumulated observation time \cite{Cognard2004}. Small slow down rates suggest that MSPs are not as magnetized as some younger pulsars \cite{New1995}. It is the reason why we have ignored magnetic deformations as the primary mass asymmetry mechanism in this work. This electromagnetic stability indicates that that gravitational radiation might dominate over magnetic dipole radiation as the dominant energy loss mechanism in MSPs.  

{\bf Crustal Strain Limit: }
There are several mechanisms that contribute to the mass asymmetry. Either due to its formation in a supernova, or due to the presence of an accretion disk, the neutron star's rotation and magnetic axis might not coincide. The enormous Lorentz forces on the crust might then contribute to the asymmetric mass distribution. Alternatively, the mass distribution could also be changed significantly due to star quakes, internal magnetic fields, instabilities induced by gravitational effects, or viscosity of the dense matter \cite{LyneBook2006}. There are also exotic theories involving superfluid turbulence in stellar cores. Estimating the maximum elastic deformation sustained by a neutron star is an active field of research, see ref. \cite{Haskell2008,Lasky2015}, and references therein for details. 

Ushomirsky et. al set limits on the maximum quadrupole moment for a NS in the presence of elastic forces, irrespective of the nature of strain on the crust \cite{Ushomirsky2000}. For standard parameters for $I_{zz}$ and breaking strain of the crust, this quadrupole moment leads to a maximum ellipticity of $\epsilon_{n}\sim6\times 10^{-7}$ for a conventional neutron star\cite{Owen2005}. This in turn can be used to evaluate limit on the gravitational wave strain amplitude,
\begin{equation} 
\label{eq:hspindown}
h_{\epsilon n}=-\frac{4G}{c^4 d} \epsilon_n I_{zz}\omega_p^2.
\end{equation}

Both the spin down and strain mechanisms put upper limits on the metric perturbation due to different physics. Therefore, we assume that the strain due to GW from pulsars is smaller than the lower of the two limits.

Table \ref{tab:LIGO_POI}  shows the observational data, along with theoretical and measured estimates of $h$ for pulsars of interest from the LIGO+VIRGO data. 
\begin{widetext}
\begin{center}
\begin{table}[ht]
\begin{center}
\begin{tabular}{|c |c| c| c| c| c| c|}
\hline
Pulsar & $\omega_p/2\pi$ & $\dot{\omega}_p/2\pi$  (Hz s$^{-1}$) & $d$ (kpc) &$h_{sd}$ & $h_{\epsilon n}$ &$h_0^{95\%}$  \\
\hline\hline

J0534+2200 (Crab)& 29.72 & $ -3.7 \times 10^{-10}$ & 2.0 & $1.4 \times 10^{-24}$ & $1.1 \times 10^{-27}$ & $1.6 \times 10^{-25}$ \\
\hline

J0537-6910 (N157B)& 61.97 & $ -2.0 \times 10^{-10}$ & 50.0 & $3.0 \times 10^{-26}$ & $1.9 \times 10^{-28}$ & $3.8 \times 10^{-26}$ \\
\hline

J0835-4510 (Vela)& 11.19 & $ -1.6 \times 10^{-11}$ & 0.29 & $3.3 \times 10^{-24}$ & $1.1 \times 10^{-27}$ & $1.1 \times 10^{-24}$ \\
\hline

J1813-1246 & 20.80 & $ -7.6 \times 10^{-12}$ & 1.9 & $2.6 \times 10^{-25}$ & $5.8 \times 10^{-28}$ & $3.4 \times 10^{-25}$ \\
\hline

J1833-1034 & 16.16 & $ -5.3 \times 10^{-11}$ & 4.8 & $3.0 \times 10^{-25}$ & $1.4 \times 10^{-28}$ & $1.3 \times 10^{-24}$ \\
\hline

J1913+1011 & 27.85 & $ -2.6 \times 10^{-12}$ & 4.5 & $2.3 \times 10^{-25}$ & $4.4 \times 10^{-28}$ & $1.6 \times 10^{-25}$ \\
\hline

J1952+3252 & 25.30 & $ -3.7 \times 10^{-12}$ & 3.0 & $1.0 \times 10^{-25}$ & $5.4 \times 10^{-28}$ & $2.7 \times 10^{-25}$ \\
\hline

\end{tabular}
\end{center}
\label{tab:LIGO_POI}
\caption{Table of Pulsars of interest from ref. \cite{Aasi2014}. $h_{\epsilon n}$ gives the strain (from Eq. \ref{eq:hmax}) if the ellipticity was $\epsilon=6\times 10^{-7}$, the maximum for a neutron star made of neutrons \cite{Owen2005}. We have assumed $I_{zz}=10^{38}$ kg-m$^2$, consistent with LIGO literature. }
\end{table}
\end{center}
\end{widetext}

It is interesting to note that for the younger pulsars with small rotation frequencies and large spin down rates, the upper limit on GW strain is set by the elastic deformation limit, while for MSPs smaller spin-down rates lead to a significantly lower limit set by $h_{sd}$. In both cases, the GW signal limit is typically below $10^{-27}$. These limits provide an upper limit on GW strength due to different physics, it is possible (in fact expected) that the actual signal would be even lower. However, it is worth remembering that the observation (or even absence) of GW signal is the only known way to gain information about the interior of these exotic objects.

\section{Search for optimal detector geometry}
As eq. \ref{eq:Hedet_hmin} suggests, the minimum detectable strain by the helium detector depends on several parameters. Thus, it is difficult to determine the best geometry for gravitational wave detection. Here we analyze several acoustic modes for a cylindrical detector that have a non-zero quadrupole tensor. We have chosen each of these geometries/modes to have a resonance frequency around 1075 Hz, and assumed a $Q$-factor of $10^{11}$. In particular, $h_{\rm min}$ is evaluated for pulsar J1843-1113. 

Before presenting a table analyzing 7 lower modes of interest, we present a summary of some general trends:
\begin{enumerate}
\item{For the same mode and frequency, it is always advantageous to use a bigger mass (assuming $Q$-factor remains the same). }
\item{Higher $n$ modes have significantly smaller effective area than lower $n$ modes. This then contributes to lower strain sensitivity. Thus, it is advantageous to have the lowest $n$ mode for a given frequency.}
\item{The maximum of the directivity function ($d^A(\theta,\phi)$) can vary by up to a factor of 4 in cylindrical geometry depending on the mode of interest.}
\end{enumerate}
Below we summarize various properties of ten different cylindrical geometries with similar resonance frequencies. Geometry 7 and 8 are used in the main text.
\begin{widetext}
\begin{center}
\begin{table}[ht]
\label{tab:Modestable}
\begin{tabular}{|c |c |c| c| c| c| c|  c| }
\hline
\#& Mode & Dimensions & Red. mass & Effective area & Quadrupole tensor (kg-m$^2$) & Directivity & $h_{\rm min} (/\sqrt{\rm Hz})$\\
\hline
1& $[001]$ & a= 0.135m, L= 0.1m 
& $\mu= 0.42 M$ & $A_G=0.14 \pi r^2$
& $q=0.028\times \begin{bmatrix}-1/2 &0 &0 \\ 0& -1/2 &0 \\0 & 0 & 1\end{bmatrix}$ & $d_{\rm max}$= 1.875 
& $h_{\rm min} = 1.69 \times 10^{-22}$\\
\hline
2 & $[001]$ & a= 0.135m, L= 0.3m 
& $\mu= 0.42 M$ & $A_G=0.14 \pi r^2$
& $q=0.084\times \begin{bmatrix}-1/2 &0 &0 \\ 0& -1/2 &0 \\0 & 0 & 1\end{bmatrix}$ & $d_{\rm max}$= 1.875 
& $h_{\rm min} = 9.78 \times 10^{-23}$\\
\hline
3 & $[002]$ & a= 0.245m, L= 0.1m 
& $\mu= 0.29 M$ & $A_G=0.01 \pi r^2$
& $q=0.01\times \begin{bmatrix}-1/2 &0 &0 \\ 0& -1/2 &0 \\0 & 0 & 1\end{bmatrix}$ & $d_{\rm max}$= 1.875 
& $h_{\rm min} = 5.99 \times 10^{-22}$\\
\hline
5 & $[020]$ & a= 0.107m, L= 0.1m 
& $\mu= 0.51 M$ & $A_G=0.63 \pi r^2$
& $q=0.028\times \begin{bmatrix}1 &0 &0 \\ 0& -1 &0 \\0 & 0 & 0\end{bmatrix}$ & $d_{\rm max}$= 2.5 ($\psi=0$) 
& $h_{\rm min} = 1.10 \times 10^{-22}$\\
\hline
6 & $[021]$ & a= 0.235m, L= 0.1m 
& $\mu= 0.34 M$ & $A_G=0.08 \pi r^2$
& $q=0.027\times \begin{bmatrix}1 &0 &0 \\ 0& -1 &0 \\0 & 0 & 0\end{bmatrix}$ & $d_{\rm max}$= 2.5 ($\psi=0$)
& $h_{\rm min} = 2.05 \times 10^{-22}$\\
\hline
7 & $[110]$ & a= 0.15m, L= 0.123m 
& $\mu= 0.14 M$ & $A_G=0.31 \pi r^2$
& $q=0.034\times \begin{bmatrix}0&0 &1 \\ 0& 0 &0 \\1 & 0 & 0\end{bmatrix}$ & $d_{\rm max}$= 2.5 
& $h_{\rm min} = 7.23 \times 10^{-23}$\\
\hline
8 & $[110]$ & a= 0.45m, L= 0.112m 
& $\mu= 0.10 M$ & $A_G=0.26 \pi r^2$
& $q=-0.66\times \begin{bmatrix}0&0 &1 \\ 0& 0 &0 \\1 & 0 & 0\end{bmatrix}$ & $d_{\rm max}$= 2.5 
& $h_{\rm min} = 9.25 \times 10^{-24}$\\
\hline
9 & $[111]$ & a= 0.22m, L= 0.21m 
& $\mu= 0.24 M$ & $A_G=0.04 \pi r^2$
& $q=-0.091\times \begin{bmatrix}0&0 &1 \\ 0& 0 &0 \\1 & 0 & 0\end{bmatrix}$ & $d_{\rm max}$= 2.5 
& $h_{\rm min} = 6.96 \times 10^{-23}$\\
\hline
10 & $[201]$ & a= 0.3m, L= 0.247m 
& $\mu= 0.28 M$ & $A_G=0.04 \pi r^2$
& $q=-0.33\times \begin{bmatrix}1/2&0 &0 \\ 0& 1/2 &0 \\0 & 0 & -1\end{bmatrix}$ & $d_{\rm max}$= 1.875 
& $h_{\rm min} = 4.12 \times 10^{-23}$\\
\hline
\end{tabular}
\caption{Table of modes and geometries of interest. We have chosen each of these geometries/modes to have a resonance frequency of $1075\pm 5$ Hz, and assumed a $Q$-factor of $10^{11}$. In particular, $h_{\rm min}$ is evaluated for pulsar J1843-1113. $\psi=\pi/2$, unless otherwise noted.}
\end{table}
\end{center}
\end{widetext}

We find that for $l=0$ modes, one needs a cylinder with length in meters to beat the 1kHz sensitivity limit of advanced LIGO. The [020] mode has a particularly strong coupling to gravity waves due to its quadrupolar mode shape, and we choose this mode for the detector geometry discussed in the main text.

Due to its large effective area and directivity, the $[110]$ mode also efficiently couples to gravitational metric strain. Unfortunately, the [110] mode does not couple to microwaves, so the gravitational wave signal in this acoustic mode cannot be detected using our proposed optomechanical technique. 

Finally, we would like to mention that while this paper deals exclusively with cylindrical geometry, there possibly are other geometries that couple more strongly to gravitational strain. Exploring different detector geometries is an interesting numerical problem that we hope to address in the future. 


\bibliographystyle{unsrt}
\bibliography{GravitySensor}

\end{document}